\documentclass[final]{IEEEtran}
\usepackage{cite}
\usepackage{amsmath,amssymb,amsfonts,amsthm}
\usepackage{algorithm}
\usepackage{algorithmic}
\usepackage{graphicx}
\usepackage{textcomp}
\usepackage{xcolor}
\usepackage{url}
\usepackage{lipsum}
\usepackage{multirow}
\usepackage[normalem]{ulem}
\usepackage{bbding}
\usepackage{pifont}
\usepackage{wasysym}
\usepackage{bbm}
\usepackage{dsfont}
\usepackage{mathtools, nccmath}
\usepackage{makecell}
\usepackage{aligned-overset}
\usepackage{stfloats}
\usepackage{float}
\usepackage{soul}
\usepackage{threeparttable}
\ifCLASSOPTIONcompsoc
\usepackage[caption=false, font=normalsize, labelfont=sf, textfont=sf]{subfig}
\else
\usepackage[caption=false, font=footnotesize]{subfig}
\fi
\newfloat{procedure}{tbp}{lop}
\floatname{procedure}{Procedure}

\DeclareMathAlphabet{\mathbcal}{OMS}{cmsy}{b}{n}
\def\BibTeX{{\rm B\kern-.05em{\sc i\kern-.025em b}\kern-.08em
		T\kern-.1667em\lower.7ex\hbox{E}\kern-.125emX}}
\renewcommand{\baselinestretch}{0.98}
\newtheorem{remark}{Remark}

\newtheorem{prop}{Proposition}
\newtheorem{theo}{Theorem}
\newtheorem{lemma}{Lemma}
\newtheorem{corollary}{Corollary}

\begin{document}
	
	\title{%
	Near-Field Communication with Massive Movable Antennas: An {Electrostatic Equilibrium} Perspective%
	}
	\author{Shicong~Liu,~\IEEEmembership{Graduate~Student~Member,~IEEE}, Xianghao~Yu,~\IEEEmembership{Senior~Member,~IEEE}, Shenghui~Song,~\IEEEmembership{Senior~Member,~IEEE}, and Khaled~B.~Letaief,~\IEEEmembership{Fellow,~IEEE}
		\thanks{
			Shicong Liu and Xianghao Yu are with the Department of Electrical Engineering, City University of Hong Kong, Hong Kong (email: sc.liu@my.cityu.edu.hk, alex.yu@cityu.edu.hk).
			
			Shenghui Song and Khaled~B.~Letaief are with the Department of Electronic and Computer Engineering, The Hong Kong University of Science and Technology, Hong Kong (email: eeshsong@ust.hk, eekhaled@ust.hk).
		}
	}
	
	\maketitle
	
	\begin{abstract}
		Recent advancements in large-scale position-reconfigurable antennas have opened up new dimensions to effectively utilize the spatial degrees of freedom (DoFs) of wireless channels. 
		However, the deployment of existing antenna placement schemes is primarily hindered by their limited scalability and frequently overlooked near-field effects in large-scale antenna systems. 
		In this paper, we propose a novel antenna placement approach tailored for near-field massive multiple-input multiple-output systems, which effectively exploits the spatial DoFs to enhance spectral efficiency. 
		For that purpose, we first reformulate the antenna placement problem in the angular domain, resulting in a weighted Fekete problem. We then derive the optimality condition and reveal that the {optimal} antenna placement is in principle 
		an electrostatic equilibrium problem. 
		To further reduce the computational complexity of numerical optimization, we propose an ordinary differential equation (ODE)-based framework to efficiently solve the equilibrium problem. 
		In particular, the optimal antenna positions are characterized by the roots of the polynomial solutions to specific ODEs in the normalized angular domain. By simply adopting a two-step eigenvalue decomposition (EVD) approach, the optimal antenna positions can be efficiently obtained. 
		Furthermore, we perform an asymptotic analysis when the antenna size tends to infinity, which yields a closed-form solution. %
		Simulation results demonstrate that the proposed scheme efficiently harnesses the spatial DoFs of near-field channels with prominent gains in spectral efficiency and maintains robustness against system parameter mismatches. In addition, the derived asymptotic closed-form {solution} closely approaches the theoretical optimum across a wide range of practical scenarios.
	\end{abstract}
	
	
	\section{Introduction}
	\bstctlcite{IEEEexample:BSTcontrol}

The transition towards the anticipated sixth-generation (6G) wireless networks is being propelled by the soaring growth of data traffic and the increasing density of service demands~\cite{10379539}. 
These trends impose increasingly stringent requirements on network capacity and coverage. To accommodate these challenges, massive multiple-input multiple-output (MIMO) technology has garnered considerable attention in the ongoing 6G research. By leveraging large-scale antenna arrays and advanced beamforming techniques~\cite{7397861}, massive MIMO significantly enhances spatial multiplexing capabilities, thereby improving spectral efficiency and extending spatial coverage.

As the aperture of massive MIMO systems increases, the near-field region expands quadratically. 
{Unfortunately, conventional arrays with uniformly spaced fixed-position elements can suffer significant performance degradation in this regime~\cite{10555049,9696209}. A key reason is that their effectiveness is inherently tied to uniformly rich-scattering environments under far-field conditions~\cite{10.5555/1111206}.} 
In the non-uniform scattering environments such as the line-of-sight (LoS)-dominant scenario that may arise in the near-field region~\cite{10078317}, the performance of uniform arrays becomes inherently constrained~\cite{10293813}. On the other hand, the spherical nature of the EM wavefront in near-field propagation induces non-uniform spatial channel variations, rendering uniformly spaced fixed-position arrays ineffective in fully exploiting the available spatial DoFs in the near-field region~\cite{10446604,10909572}. %

Recent advancements in mechanical innovation and antenna design catalyze the development of position-reconfigurable antenna (PRA) technologies, such as the movable antenna (MA)~\cite{10286328}, fluid antenna systems (FAS)~\cite{9264694}, and the pinching antenna~\cite{11169486}. In particular, the MA-based architecture achieves spatial reconfigurability by physically repositioning antenna elements or sub-arrays~\cite{10851455,10286328}, whereas fluid antenna systems realize dynamic port selection through liquid metals~\cite{9264694} or reconfigurable radio frequency (RF) pixels~\cite{9531465}. The pinching antenna, on the other hand, activates desired radiation points along a dielectric waveguide by applying localized mechanical pinching~\cite{11169486}. 
By enabling more favorable propagation conditions and precise manipulation of EM wave radiation, these technologies offer substantial improvements over the conventional fixed-position antenna arrays in overall communication performance. With the widespread adoption of massive MIMO, there is an urgent demand for optimal design strategies specifically tailored to massive PRA systems operating in the near-field region.

\subsection{Related Works}

PRAs can be {leveraged to improve} various communication performance metrics, including spectral efficiency (SE)~\cite{10354003,10504625,10508218,10293813,10909572,10794752}, sensing performance bounds~\cite{10839251,10643473,10476966,10446604}, and secrecy rates~\cite{10092780,10416363,10901621}, where substantial performance gains have been demonstrated. To improve the SE performance of PRA-assisted wireless systems, antenna position placement strategies based on numerical optimization approaches have been extensively studied. In particular, 
a transmit power minimization method subject to individual rate constraints was proposed in~\cite{10354003}, where the authors developed a multi-directional descent framework to search multiple candidate gradient directions in parallel. 
{Moreover, the highly non-convex capacity maximization problem for FAS-assisted systems was studied in~\cite{10794752}, where the transmit covariance matrix at the base station side and antenna positions at the user side were jointly optimized by alternating optimization algorithms.} 
Besides, the weighted sum-rate maximization problem of multiuser MA system was addressed in~\cite{10504625}, where the antenna placement problem was reformulated into a weighted minimum mean square error form for tractable optimization. Building on this, 
a block coordinate descent-based algorithm incorporating practical movement constraints was further developed to reduce the computational complexity by $30\%$ without incurring notable performance degradation. 
In addition, the downlink receive power maximization problem was investigated~\cite{10508218} with a single-antenna user, where the antenna placement problem is transformed into discrete sampling point selection. Leveraging graph-theoretic tools, the original problem was recast into a solvable form, and an algorithm was accordingly developed to solve it in polynomial time. 

However, the aforementioned antenna placement strategies did not take near-field effect into consideration, which becomes non-negligible when the antenna movement range extends to hundreds of wavelengths~\cite{10909572}. 
Recognizing this limitation, a few recent studies have attempted to incorporate near-field effect into PRA system design. Specifically, to overcome the sophisticated properties of near-field channels, particle swarm optimization has been adopted as a heuristic approach for antenna position design under near-field channel models~\cite{10742118}. Besides, an alternating optimization method has also been developed to iteratively optimize the antenna positions, analog beamforming vectors, and power allocation in multiuser scenarios with multiple fixed-spacing movable sub-arrays~\cite{10909572}. {Furthermore, to maximize the energy efficiency, an alternating optimization algorithm that iteratively optimizes the beamforming and antenna position subproblems was proposed for near-field FA systems~\cite{10767351}.}

\subsection{Motivations}

Despite existing performance improvements in SE, sensing accuracy, and security achieved by dedicated antenna position designs, several important limitations remain to be addressed. First of all, most existing approaches rely on numerical optimization algorithms~\cite{10909572,10354003,10504625,10643473,10839251,10416363}, which typically involve recursively nested iterative procedures. Consequently, the computational complexity of these methods scales polynomially 
with respect to the number of antenna elements, restricting their applicability to systems with only a moderate number of antennas~\cite{10354003,10504625,10508218,10839251,10416363}. %

Second, numerical optimization algorithms often provide limited physical insight. Prior studies typically formulate antenna placement as a purely numerical optimization problem, which obscures the underlying {physical} mechanisms behind the obtained solutions, and offers little guidance for systematic array geometry design. In particular, these approaches generally fail to yield closed-form expressions for optimal antenna positions or to inform the selection of key system parameters.

Lastly, although the vast movable range significantly enlarges the effective array aperture, which thus extends the near-field region, current studies fail to incorporate customized designs tailored for the near field~\cite{10742118,10909572,10767351}. 
In particular, most existing approaches directly migrated classic design methodologies for far-field systems to their near-field counterparts. However, when the near-field effect becomes prominent, the available spatial DoFs of the wireless channel increase substantially~\cite{10945401,10909572,10742118}, which in turn calls for dedicated antenna position design to fully harness these additional DoFs. In a word, to the best of the authors' knowledge, there still lacks a low-complexity yet effective design framework for PRA-based near-field massive MIMO systems. %

\subsection{Contributions}
In this paper, we investigate the optimal antenna placement scheme to maximize SE by fully utilizing the spatial DoFs of wireless channels in point-to-point near-field communication. 
Our contributions are summarized as follows.
\begin{itemize}
	\item To facilitate near-field analysis and decouple the impacts of the transmit and receive antenna positions on SE performance, we reformulate the SE maximization problem by introducing a trigonometric substitution. The reformulated problem then reduces to a Vandermonde determinant maximization problem, also referred to as the weighted Fekete problem, which is proven to be convex.
	\item We then reveal the close connection between the first-order optimality conditions of the weighted Fekete problem and the electrostatic equilibrium model with three fixed charges. In this model, the optimal antenna coordinates exactly correspond to the equilibrium positions of multiple movable electric charges in the angular domain, under mutual repulsive forces that are inversely proportional to their separation distances.
	\item To alleviate the computational complexity of numerically solving the electrostatic equilibrium problem, polynomial theory is leveraged to characterize the equilibrium points as the roots of a monic polynomial, based on which a second-order ordinary differential equation (ODE)-based framework is further developed. With Sturm-Liouville theory, a low-complexity algorithm is further developed, which only requires a two-step eigenvalue decomposition (EVD) operation, thereby yielding a closed-form solution to the optimal antenna positions.
	\item We finally conduct asymptotic analysis of the proposed ODE-based framework when {the number of antennas on the transmitting array tends} to infinity, and derive an asymptotic closed-form solution under practical near-field conditions. The solution reveals that the optimal antenna distribution in near-field massive MIMO systems follows an arcsine distribution, establishing a fundamental connection between SE maximization and equilibrium measures of antenna positions.
\end{itemize}
\par {\it Notations}: We use normal-face, lowercase boldface, and uppercase boldface letters to denote scalars, vectors, and matrices, respectively. 
The element at the $k$-th row and the $m$-th column of matrix $\bf H$ is denoted as ${\bf H}[k,m]$, and the $n$-th element in the vector $\bf h$ is denoted by ${\bf h}[n]$. $\{{\bf H}_n\}_{n=1}^N$ represents a matrix set with the cardinality of $N$. The superscripts $(\cdot)^{T}$ and $(\cdot)^{H}$ represent the transpose and conjugate transpose operators, respectively, and $\det(\cdot)$ denotes the determinant operator. 
The imaginary unit is represented as $\jmath$ such that $\jmath^2=-1$. $\mathbb{R}$ and $\mathbb{Z}$ denote the sets of real numbers and integers, respectively. We denote by $f^{\prime}(x)$ and $f^{\prime\prime}(x)$ the first and second derivatives of a continuously differentiable function $f(x)$, respectively. 

\section{System Model}

In this section, we present the system model for antenna position design, including the massive movable antenna configuration and the channel model.
\subsection{Massive Position-Reconfigurable Antennas}
\begin{figure}
	\centering
	\vspace{-3mm}
	\includegraphics[width=0.425\textwidth]{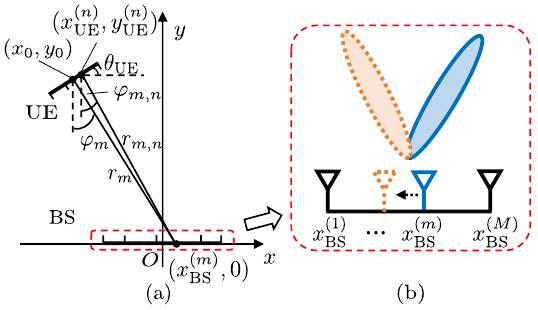}
	\vspace{-3mm}
	\caption{(a) The considered near-field communication scenario under the Cartesian coordinate system, and (b) the schematic diagram of massive movable antennas, with the $1$-st and $M$-th antenna fixed at the edges.}
	\label{fig:sysmodel}
	\vspace{-3mm}
\end{figure}

Consider a point-to-point massive MIMO downlink transmission scenario with $M$ movable antennas\footnote{To simplify the problem and promote practicality, in this paper, we only consider movable antennas at the BS, with fixed antenna elements at the UE side. However, due to the spatial duality of the channel model, the proposed methodology in this paper can also be extended to UEs with movable antennas.} at the base station (BS) and $N$ fixed-position antennas at the user equipment (UE). As illustrated in Fig.~\ref{fig:sysmodel}, linear arrays are considered at transceivers. The Cartesian coordinate system is established such that the centroid of the BS antenna array is located at the origin $O(0,0)$ and the BS array aligns along the $x$-axis. The coordinates of the $m$-th element on the BS array and the $n$-th element on the UE array are given by $(x_{\rm BS}^{(m)},0)$ and $(x_{\rm UE}^{(n)},y_{\rm UE}^{(n)})$, respectively. 

In this paper, the $x$ coordinates of BS antenna elements, i.e., $\{ x_{\rm BS}^{(1)}, x_{\rm BS}^{(2)},\cdots,x_{\rm BS}^{(M)} \}$, are in the ascending order specified by
\begin{equation}
	-\frac{A_{\rm T}}{2}=x_{\rm BS}^{(1)} < x_{\rm BS}^{(2)} <\cdots < x_{\rm BS}^{(M)} = \frac{A_{\rm T}}{2},
	\label{eq:xseq}
\end{equation}
where $A_{\rm T} = (M-1)d$ is the aperture of BS array, and $d$ denotes the unit antenna spacing. Specifically, for the conventional uniform linear arrays (ULAs), the antenna positions can be given by 
\begin{equation}
	x_{\rm BS,ULA}^{(m)} = -\frac{(M-1)d}{2} + (m-1)d,\label{eq:uni}
\end{equation}
where $m \in \{ k\in \mathbb{Z} \mid 1\leq k\leq M \}$. 
\subsection{Channel Model}

In this paper, we adopt the spherical wave model~\cite{9693928} to characterize the near-field wireless channel. Specifically, the spatial channel response between the $m$-th antenna on the BS array and the $n$-th antenna on the UE array is modeled by
\begin{align}
	h\left(r_{m,n}\right) = \frac{e^{\jmath \kappa r_{m,n} }}{r_{m,n}},
	\label{eq:greenf}
\end{align}
where $\kappa = 2\pi/\lambda$ is the wavenumber with $\lambda$ being the carrier wavelength. In addition, $r_{m,n}$ denotes the distance between the $m$-th antenna at the BS and the $n$-th antenna at the UE, given by
\begin{equation}
	r_{m,n}= \sqrt{\left( y_{\rm UE}^{(n)} \right)^2 + \left( x_{\rm UE}^{(n)} - x_{\rm BS}^{(m)} \right)^2}.
	\label{eq:rmn}
\end{equation}

Since the LoS path has significantly higher signal strength than the non-line-of-sight (NLoS) counterparts in the near-field region~\cite{10571546}, the $(n,m)$-th element of the channel matrix ${\bf H}_{\bf x}$ is predominantly determined by the LoS path given by
\begin{align}
	{\bf H}_{{\bf x}}[n,m]
	= h\left(r_{m,n}\right). %
	\label{eq:chmodel_los}
\end{align}
In~\eqref{eq:chmodel_los}, ${\bf x}= [x_{\rm BS}^{(1)}, x_{\rm BS}^{(2)},\cdots,x_{\rm BS}^{(M)}]$ denotes the positions of the BS antenna elements on the $x$-axis, which are involved in the distance terms $r_{m,n}$. 
Hence, the downlink channel matrix is a function of ${\bf x}$.

In this paper, we assume that the channel state information (CSI) is perfectly known at the transmitter, as it can be acquired using advanced estimation methods~\cite{10845870,10643473,10.1007/978-3-540-88793-5_2}. {Nevertheless, the proposed scheme remains robust under parameter mismatches, as demonstrated in the numerical results.}

\section{Problem Formulation and Analysis}
\label{sec:formulation}
In this section, we reformulate the near-field SE maximization problem in the angular domain, which simplifies the derivation and decouples the impact of the transmit and receive antenna positions. The SE maximization problem is then recast as a \textit{weighted Fekete problem}.

\subsection{Problem Statement}
With a given antenna position vector ${\bf x}$, SE is defined as
\begin{equation}
	\begin{aligned}
		R_{\bf x}=\log_2\det \left( {\bf I}_N + \frac{1}{ \sigma_{\rm n}^2} {\bf H}_{{\bf x}}^{} {\bf Q} {\bf H}_{{\bf x}}^{H} \right),
		\label{eq:cap}
	\end{aligned}
\end{equation}
where $\sigma_n^2$ is the power of additive white Gaussian noise (AWGN), and ${\bf Q}$ denotes the transmit covariance matrix. Since the primary focus of this paper is on the performance gains introduced by the MA, we assume an isotropic transmission with ${\bf Q} = \frac{P_{\rm T}}{M} {\bf I}_{M}$, where $P_{\rm T}$ denotes the total transmit power. Under such circumstances, the SE can be further given by
\begin{equation}
	R_{\bf x} = \log_2\det\left( {\bf I}_N + \frac{\rho}{M}{\bf H}^{}_{{\bf x}} {\bf H}_{{\bf x}}^{H}  \right),
\end{equation}
where $\rho = \frac{P_{\rm T}}{\sigma_{\rm n}^2}$
denotes the signal-to-noise ratio (SNR). %
Our objective is to determine the optimal antenna positions ${\bf x}$ such that, by reorganizing the intrinsic structure of the channel matrix ${\bf H}_{\bf x}$, more orthogonal transmission modes can be excited and exploited to achieve a higher SE. Accordingly, the rate maximization problem is formulated as
\begin{equation}
	\mathcal{P}_1:~~
	\begin{aligned}
		&\underset{{\bf x}}{\max} && {R}_{\bf x}\\[-4pt]
		&\mathrm{s.t.}&&{\bf x}[M]=-{\bf x}[1] = \frac{A}{2},\\
		& ~ &&{\bf x}[m+1]-{\bf x}[m]>0,\quad1\leq m \leq M-1,
	\end{aligned}
	\label{eq:P1}
\end{equation}
where the first constraint confines the moving area within the aperture $[-A/2,A/2]$, while the second enforces the ordered structure of ${\bf x}$ in~\eqref{eq:xseq}.

In the near-field region, the received signal exhibits a much slighter fading effect compared to the far-field region, and hence, the high SNR condition $\rho\gg 1$ is more likely to hold given a power budget. The original SE objective in~\eqref{eq:cap} can then be accurately approximated by 
\begin{equation}
	\begin{aligned}
		{R}_{\bf x} \simeq %
		\log_2 \det \left( {\rho} {\bf H}_{{\bf x}}^{} {\bf H}_{{\bf x}}^H \right)\triangleq\tilde{R}_{\bf x}.
		\label{eq:overall_obj}
	\end{aligned}
\end{equation}
Revisiting the channel model in~\eqref{eq:chmodel_los}, we next derive the distance terms $r_{m,n}$ required for ${\bf H}_{\bf x}$ in~\eqref{eq:overall_obj} to accomplish the formulation. 
As shown in Fig.~\ref{fig:sysmodel}, the centroid coordinate of the UE array is $(x_0,y_0)$, and the coordinate of the $n$-th antenna on the UE array is given by
\begin{equation}
	\begin{cases}
		x_{\rm UE}^{(n)} &= x_0 + d_n \cos\theta_{\rm UE}\\
		y_{\rm UE}^{(n)} &= y_0 + d_n \sin\theta_{\rm UE}
	\end{cases},
	\label{eq:xnyn}
\end{equation}
where $\theta_{\rm UE}$ represents the azimuth angle of the UE array, and $d_n$ denotes the relative distance from the $n$-th element to the centroid $(x_0,y_0)$. By substituting~\eqref{eq:xnyn} into~\eqref{eq:rmn}, we have
\begin{align}
	&r_{m,n}\label{eq:rmn2}\\
	={}& \sqrt{\left( y_0 \!+\! d_n\sin\theta_{\rm UE} \right)^2 \!+\! \left( x_0 + d_n\cos\theta_{\rm UE} \!-\!x_{\rm BS}^{(m)}\right)^2}\notag\\
	={}&\sqrt{r_m^2+2d_n\!\left( y_0\sin\theta_{\rm UE}\!+\!\left( x_0\!-\!x_{\rm BS}^{(m)} \right) \cos\theta_{\rm UE}  \right)+d_n^2  }\notag
\end{align}
where $r_m^2 = y_0^2 + ( x_0-x_{\rm BS}^{(m)} )^2$ represents the squared distance from the centroid of the UE to the $m$-th element on the BS array. 
In a typical near-field communication scenario, the communication distance $r_{m}$ is far larger than the aperture of the UE array, i.e., $r_m\gg \max_n (d_n)$. Therefore, the second-order small quantity $d_n^2$ in~\eqref{eq:rmn2} is negligible and we perform a binomial expansion as 
\begin{equation}
	\begin{aligned}
		r_{m,n} &\simeq r_m \!+\! \frac{d_n}{r_m}\left(  y_0 \sin\theta_{\rm UE} + \left( x_0-x_{\rm BS}^{(m)} \right) \cos\theta_{\rm UE} \right)\\
		&\triangleq \tilde{r}_{m,n}.
	\end{aligned}
	\label{eq:rmn3}
\end{equation}
According to Fig.~\ref{fig:sysmodel}, we have $\cos\varphi_m=y_0/r_m$ and $\sin\varphi_m=(x_\mathrm{BS}^{(m)}-x_0)/r_m$, and therefore~\eqref{eq:rmn3} further yields
\begin{equation}
	\begin{aligned}
		\tilde{r}_{m,n} &= r_m + d_n\left( \sin\theta_{\rm UE} \cos\varphi_m- \cos\theta_{\rm UE}\sin\varphi_m\right)\\
		& = y_0\sec\varphi_m + d_n \sin\left( \theta_{\rm UE} - \varphi_m \right).
	\end{aligned}\label{eq:tri}
\end{equation}

Note that in the spherical wave model~\eqref{eq:greenf}, the phase term in the numerator has a significantly greater impact on the channel characteristics than the distance term in the denominator. This is because the wavenumber $\kappa = 2\pi/\lambda$ is typically very large, especially for high carrier frequencies. In contrast, the physical size of the UE array is generally small, resulting in only limited absolute variation in distance $\tilde{r}_{m,n}$ across different UE antennas. To facilitate tractable analysis, we therefore neglect the $d_n$-dependent variation in $\tilde{r}_{m,n}$, and retain only $r_m$ in the denominator of~\eqref{eq:greenf}. Substituting the distance terms $\tilde{r}_{m,n}$ and $r_m$ into the phase and denominator of the channel model~\eqref{eq:greenf}, respectively, we have
\begin{equation}
	\begin{aligned}
		{\bf H}_{{\bf x}}^{}[n,m] \simeq \frac{e^{\jmath\kappa \left( r_m + d_n\sin\left( \theta_{\rm UE}-\varphi_m \right) \right) }}{r_m} %
		\triangleq \tilde{\bf H}_{{\bf x}}[n,m].
	\end{aligned}
	\label{eq:approxch}
\end{equation}

\subsection{Weighted Fekete Problem Formulation}
Since the denominator $r_m$ depends only on the antenna index $m$ of the BS array, the approximated channel matrix $\tilde{\bf H}_{{\bf x}}$ can be decomposed by
\begin{equation}
	\tilde{\bf H}_{{\bf x}} = {\bf P} {\bf D}_{\rm T},%
	\label{eq:HPD}
\end{equation}
where 
${\bf D}_{\rm T}$ is an $M$-dimensional diagonal matrix given by
\begin{equation}
	{\bf D}_{\rm T}={\rm diag}\left(\left[ \frac{e^{\jmath \kappa r_1}}{r_1},\frac{e^{\jmath \kappa r_2}}{r_2},\cdots,\frac{e^{\jmath \kappa r_M}}{r_M} \right]  \right),
\end{equation}
and $\bf P$ contains only the phase terms with cross-dependence between $m$ and $n$.  
We then employ Taylor expansion~\cite{7546944} for its $(n,m)$-th entry as
\begin{equation}
	\begin{aligned}
		{\bf P}[n,m] &= e^{-\jmath\kappa d_n\sin\left(\varphi_m - \theta_{\rm UE}\right) }\\ 
		&= \sum_{k=0}^{\infty} \frac{\left( -\jmath \kappa d_n\sin\left(\varphi_m - \theta_{\rm UE}\right) \right)^k}{k!}.
	\end{aligned}\label{eq:pwang}
\end{equation}
Hence, matrix $\bf P$ can be decomposed as
\begin{equation}
	{\bf P} = {\bf V}_{\rm R} \boldsymbol{\Sigma} {\bf V}_{\rm T}^H,
	\label{eq:decomp_vander_}
\end{equation}
with $\boldsymbol{\Sigma} = {\rm diag}\left( (-\jmath )^0/0!, (-\jmath )^1/1!,\cdots\right)$ being a diagonal matrix with {infinite} columns, while ${\bf V}_{\rm T}$ and ${\bf V}_{\rm R}$ are Vandermonde matrices with infinite columns %
sharing a similar form as
\begin{equation}
	\renewcommand{\arraystretch}{0.7}
	{\bf V}_{\rm T} = \left[\begin{matrix}
		1 & \sin\left(\varphi_1 - \theta_{\rm UE}\right) & \sin^2\left(\varphi_1 - \theta_{\rm UE}\right) &\cdots\\
		1 & \sin\left(\varphi_2 - \theta_{\rm UE}\right) & \sin^2\left(\varphi_2 - \theta_{\rm UE}\right) &\cdots\\
		\vdots &\vdots & \vdots&\cdots\\
		1 & \sin\left(\varphi_M - \theta_{\rm UE}\right) & \sin^2\left(\varphi_M - \theta_{\rm UE}\right) &\cdots
	\end{matrix}  \right]
	\label{eq:vandermonde_t}
\end{equation}
and
\begin{equation}
	\renewcommand{\arraystretch}{0.7}
	{\bf V}_{\rm R} = \left[\begin{matrix}
		1 & \kappa d_1 & \left( \kappa d_1 \right)^2&\cdots\\
		1 & \kappa d_2 & \left( \kappa d_2 \right)^2&\cdots\\
		\vdots &\vdots & \vdots&\cdots\\
		1 & \kappa d_N & \left( \kappa d_N \right)^2 &\cdots
	\end{matrix}  \right].
	\label{eq:vandermonde_r}
\end{equation}

Note that the absolute values of the diagonal entries in $\boldsymbol{\Sigma}$
\begin{equation}
	\left\vert \frac{\left( -\jmath \right)^k}{k!}\right\vert  = \frac{1}{k!}%
\end{equation}
decay factorially with the value of $k$. Therefore, we can truncate the matrices $\boldsymbol{\Sigma}$, ${\bf V}_{\rm T}$, and ${\bf V}_{\rm R}$ to $M$ columns, with negligible error (when $M$ is large) as
\begin{equation}
	{\bf P} \simeq \uline{\bf V}_{\rm R} \uline{\boldsymbol{\Sigma}}\,  \uline{\bf V}_{\rm T}^H \triangleq \uline{\bf P}_M.
	\label{eq:truncate_}
\end{equation}
Substituting~\eqref{eq:approxch}, \eqref{eq:HPD}, and~\eqref{eq:truncate_} into~\eqref{eq:overall_obj}, we further simplify the objective function $\tilde{R}_{\bf x}$ in the near-field region as
\begingroup
\allowdisplaybreaks
\begin{align}
	& \tilde{R}_{\bf x} = \log_2 \det \left( {\rho} {\bf H}_{{\bf x}}^{} {\bf H}_{{\bf x}}^H \right)\notag\\
	\simeq{}& N\log_2\rho + \log_2\det\left( {\bf D}_{\rm T}^{}  {\bf D}_{\rm T}^H \right)+\log_2\det\left(\underline{\bf P}_M^{~} \underline{\bf P}_M^H\right)  \\
	={}& R_0+\log_2\det\left( {\bf D}_{\rm T}^{} {\bf D}_{\rm T}^H \right) + \log_2 \det  \left( \underline{\bf V}_{\rm T}^H \underline{\bf V}_{\rm T} \right),\notag
\end{align}
\endgroup
where the rate constant
\begin{equation}
	R_0=N\log_2\rho + \log_2\det\left( \underline{\bf V}_{\rm R}^{H}  \underline{\bf V}_{\rm R} \right) + \log_2 \det \left( \underline{\boldsymbol{\Sigma}}^H\underline{\boldsymbol{\Sigma}}  \right) \notag
\end{equation}
only depends on the SNR $\rho$ and the fixed geometry of the UE array. Note that by introducing the angular domain transform in~\eqref{eq:tri}, near-field approximation in~\eqref{eq:approxch}, and Taylor expansion in~\eqref{eq:pwang}, the impact of the transmitter and receiver is decoupled into {$\{{\bf D}_{\rm T}, {\bf V}_{\rm T}\}$ and ${\bf V}_{\rm R}$}, respectively. 
Therefore, maximizing the SE~\eqref{eq:overall_obj} in the near-field region is equivalent to maximizing the determinant related to the diagonal matrix ${\bf D}_{\rm T}$ and the Vandermonde matrix $\underline{\bf V}_{\rm T}^{(M)}$, which are respectively given by
\begin{equation}
	\det  \left( {\bf D}_{\rm T}^{} {\bf D}_{\rm T}^H\right) = \prod_{m=1}^{M} \frac{1}{r_m^2},
	\label{eq:logdet_weighting}
\end{equation}
and
\begin{equation}
	\det  \left(  \underline{\bf V}_{\rm T}^{H} \underline{\bf V}_{\rm T} \right)= \prod_{1\leq i< j\leq M} \left( s_j -s_i \right)^2,
	\label{eq:logdet_vandermonde}
\end{equation}
where $s_i \triangleq \sin\left( \varphi_i - \theta_{\rm UE} \right)$, and the angle $\varphi_i - \theta_{\rm UE}$ is in the range\footnote{The range is chosen based on practical considerations, since UE arrays are typically oriented so that their main boresight approximately faces the serving BS array.} $\left[ -\pi/2,\pi/2 \right]$. 
We then define the antenna position in the \textit{angular domain} as
\begin{equation}
	\begin{aligned}
		s_m &=  \sin\left( \varphi_m - \theta_{\rm UE} \right)\\
		&= \sin\left( \arctan \left( \tfrac{x_{\rm BS}^{(m)}-x_0}{y_0}\right)-\theta_{\rm UE} \right).
	\end{aligned}
	\label{eq:sdomain}
\end{equation}
In this way, the distance term $r_m$ in~\eqref{eq:logdet_weighting} is expressed by
\begin{equation}
	r_m^2 = \frac{y_0^2}{\cos^2\varphi_m} =  \frac{y_0^2}{1-\tilde{s}_m^2},
	\label{eq:weightingfunc1}
\end{equation}
where $\tilde{s}_m = \sin \varphi_m = \sin\left( \arcsin s_m  + \theta_{\rm UE} \right)$. 
By substituting~\eqref{eq:weightingfunc1} into~\eqref{eq:logdet_weighting}, we can reformulate the problem $\mathcal{P}_1$ as
\begin{equation}
	\mathcal{P}_2:
	\begin{aligned}
		&\underset{\bf s}{\max} && J({\bf s})\\[-2pt]
		&\mathrm{s.t.}&&s_{\rm min} = s_1 < s_2 <\cdots < s_M={s}_{\rm max},%
	\end{aligned}
	\label{eq:P2}
\end{equation}
where ${\bf s} = [s_1,s_2,\cdots, s_M]^T$ and
\begin{equation}
	J({\bf s}) = 2\!\!\sum_{1\leq i< j\leq M}\!\! \log_2 \left( s_j -s_i \right) + \sum_{m=1}^{M} \log_2 \left( 1-\tilde{s}_m^2 \right).\label{eq:js}
\end{equation}
{The parameters} $s_{\rm min}$ and $s_{\rm max}$ are {determined} by setting $x_\mathrm{BS}^{(m)}$ in~\eqref{eq:sdomain} as $-A/2$ and $A/2$, respectively. 
Problem $\mathcal{P}_2$ is known as the \textit{weighted Fekete problem}~\cite[p.~114, Ch. III]{Logarithmic_Potentials} with weighting function
\begin{equation}
	w(s) = 1-\tilde{s}^2=\cos^2\left( \arcsin s +\theta_{\rm UE} \right),\label{eq:weighting}
\end{equation}
which generalizes the classical Fekete problem by incorporating a position-dependent weighting function.
\begin{remark}
	In classical potential theory, the Fekete problem seeks the optimal placement of $M$ electric charges on a compact set to maximize logarithmic potential energy~\cite{Logarithmic_Potentials}. 
	In the presence of an external electric field, this framework extends to the weighted Fekete problem, where the optimal distribution of $M$ charges achieves a 
	balance between \textbf{internal} mutual repulsion and \textbf{external} influence.
	In the context of near-field communication with movable antennas, the mutual coupling among antennas corresponds to the internal interaction, while the field induced by the UE serves as the external field. Accordingly, the positions of $M$ movable antennas have to be delicately optimized to obtain a state of equilibrium between these two forces.
\end{remark}

\section{Proposed Optimal Antenna Placement Scheme}
In this section, we first analyze the convexity of problem $\mathcal{P}_2$, which guarantees the uniqueness of the optimal solution and enables its numerical computation using existing mathematical tools. To further reduce the computational complexity introduced by the intrinsic iterative procedures involved in the optimization algorithm, we further propose an ordinary differential equation (ODE)-based framework to solve the optimal antenna placement problem efficiently.

\subsection{Optimal Solution}
We first reveal the convexity of problem $\mathcal{P}_2$ by the following lemma.
\begin{lemma}
	\label{lemma:concave}
	Problem $\mathcal{P}_2$ is a concave problem with a unique optimal solution of position set $\{ s_m \}_{m=1}^M$.
\end{lemma}
\begin{IEEEproof}
	See Appendix~\ref{append:1}.
\end{IEEEproof}

Lemma~\ref{lemma:concave} guarantees the existence and uniqueness of the global maximum, which can be obtained numerically by existing toolboxes such as CVX~\cite{cvx}.

\subsection{Optimality Conditions in the Normalized Angular Domain}
\label{sec:heun}
However, numerically solving $\mathcal{P}_2$ using existing general toolboxes not only incurs excessively high computational complexity, but also offers limited physical insights into the optimal antenna placement. Therefore, in this subsection, we first reveal the close connection between the first order optimality condition and the electrostatic equilibrium model, and further propose an efficient ODE-based framework for obtaining {closed-form} solutions for the optimal antenna positions.

\begin{figure}[t]
	\centering
	\includegraphics[width=0.35\textwidth]{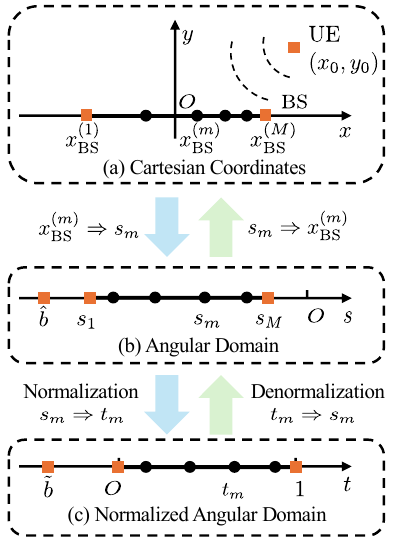}
	\vspace{-2mm}
	\caption{The relationships among antenna positions $\{ x_{\rm BS}^{m} \}_{m=1}^{M}$, the angular domain position $\{ s_m \}_{m=1}^M$, and the normalized angular domain positions $\{ t_m \}_{m=1}^M$. 
	The near-field antenna position design problem can be transformed to an electrostatic equilibrium problem with three fixed charges (marked by squares) and $M-1$ free charges (marked by disks).}
	\label{fig:eqeq}
	\vspace{-3mm}
\end{figure}

For that purpose, we first derive the first-order optimality condition of problem $\mathcal{P}_2$ as
\begin{align}
	\frac{\partial J}{\partial s_m} = \frac{1}{\ln 2}\left( \sum_{\substack{i=1\\i\neq m}}^{M} \frac{2}{s_m-s_i} - \frac{2\tan\varphi_m}{\sqrt{1-s_m^2}} \right)=0.\label{eq:firstorder_ori}
\end{align}
Since the derivative of weighting function $w(s)$ incorporates sophisticated non-linear operations, we perform a Taylor expansion of $w(s)$ at point $s_0= \sin( \varphi_0 - \theta_{\rm UE} )$ as
\begin{equation}
	\cos^2\left( \arcsin s +\theta_{\rm UE} \right) \simeq a-b(s-s_0)+ \mathcal{O}\left( \left(s-s_0\right)^2 \right),\label{eq:taylor1}
\end{equation}
where $a=\cos^2\varphi_0$, $b=\frac{\sin2\varphi_0}{\cos\left(\varphi_0 - \theta_{\rm UE}\right)}$, and $\tan \varphi_0=-{x_0}/{y_0}$. 
Therefore, the first-order optimality condition, also known as the equilibrium equation, of problem $\mathcal{P}_2$ is approximated by 
\begin{align}
	\frac{\partial J}{\partial s_m} &\simeq\frac{1}{\ln 2}\left( \sum_{\substack{i=1\\i\neq m}}^{M} \frac{2}{s_m-s_i} - \frac{b}{a-b\left(s_m-s_0\right)} \right)=0,\label{eq:firstorder}
\end{align}
where $m \in \{ k\in \mathbb{Z}\mid 2\leq k\leq M-1 \}$. Then, \eqref{eq:firstorder} further yields
\begin{equation}
	\sum_{\substack{i=2\\i\neq m}}^{M-1}\!\!\frac{2}{s_m-s_i}\!=\!-\frac{2}{s_m-s_{\rm min}}\!-\!\frac{2}{s_m-s_{\rm max}}\!-\frac{1}{s_m-\hat{b}},
	\label{eq:equilibrium}
\end{equation}
where $\hat{b} = s_0 + \frac{a}{b}$. Note that \eqref{eq:equilibrium} models an electrostatic equilibrium problem, in which three charges are fixed at positions $ s_1=s_{\min}$, $s_M = s_{\max}$, and $\hat{b}$, while the remaining $M-2$ free charges reach equilibrium under mutual repulsive forces that are inversely proportional to their separation distances, as illustrated in Fig.~\ref{fig:eqeq}(b). 
By further introducing normalization 
\begin{equation}
	t = \frac{s-s_{\rm min}}{s_{\rm max}-s_{\rm min}}\in (0,1)\label{eq:tdomain}
\end{equation}
into~\eqref{eq:equilibrium}, we finally have
\begin{equation}
	\sum_{\substack{i=2\\i\neq m}}^{M-1}\frac{2}{t_m-t_i} = - \frac{2}{t_m}-\frac{2}{t_m-1}-\frac{1}{t_m-\tilde{b}},
	\label{eq:equilibrium_norm}
\end{equation}
with 
\begin{equation}
	\tilde{b} = \frac{\hat{b}-s_{\rm min}}{s_{\rm max}-s_{\rm min}}.\label{eq:bpprime}
\end{equation} 
In this way, the electrostatic equilibrium problem described by~\eqref{eq:equilibrium} is mapped to the normalized angular domain, with three fixed charges at $0$, $1$, and $\tilde{b}$, as illustrated in Fig.~\ref{fig:eqeq}(c).

\begin{figure*}[b]
	\vspace{-3mm}
	\centering
	\rule{\textwidth}{0.5pt}%
	\begin{equation}
		p^{\prime\prime}(t)+\left( \frac{2}{t}+\frac{2}{t-1}+\frac{1}{t-\tilde{b}} \right)p^{\prime}(t)+\frac{v_1 t+v_0}{t(t-1)(t-\tilde{b})}p(t)=0\label{eq:heun}\tag{47}
	\end{equation}
\end{figure*}
\subsection{Proposed ODE-based Framework}
To cope with the normalized equilibrium equation in~\eqref{eq:equilibrium_norm}, we first construct a \textit{monic} polynomial $p(t)$ of degree $M-2$ as
\begin{equation}
	p(t) = \prod_{m=2}^{M-1}\left( t- t_m\right) = \sum_{m=0}^{M-2} c_m t^m,
	\label{eq:defp}
\end{equation}
{where the solutions $\{ t_m \}_{m=2}^{M-1}$ to the original equilibrium problem~\eqref{eq:equilibrium_norm} are now the roots of polynomial $p(t)$}, and $c_m$ denotes the coefficient of $t^m$ with $c_{M-2} = 1$. By substituting the widely used identity for polynomial roots~\cite{StieltjesPolynomials}
\begin{equation}
	\frac{p^{\prime\prime}(t_m)}{p^{\prime}(t_m)} = \sum_{\substack{i=2\\i\neq m}}^{M-1} \frac{2}{t_m-t_i}
	\label{eq:pppdpp}
\end{equation}
into~\eqref{eq:equilibrium_norm}, we obtain
\begin{equation}
	p^{\prime\prime}(t_m) =\left( -\frac{2}{t_m}-\frac{2}{t_m-1}-\frac{1}{t_m-\tilde{b}} \right)p^{\prime}(t_m).
	\label{eq:equilibrium_2}
\end{equation}
To eliminate the potential negative orders of $t_m$, we rewrite~\eqref{eq:equilibrium_2} as
\begin{align}
	&t_m(t_m-1)(t_m-\tilde{b})p^{\prime\prime}(t_m)~+\label{eq:equilibrium_temp}\\
	&\left( 2(t_m\!-\!1)(t_m\!-\!\tilde{b}) \!+\! 2t_m(t_m\!-\!\tilde{b})\!+\!t_m(t_m\!-\!1) \right)\! p^{\prime}(t_m)\!=\! 0,\notag
\end{align}
which indicates that $\{ t_m \}_{m=2}^{M-1}$ are the roots of the following polynomial $r(t)$ of degree $M-1$ as
\begin{align}
	r(t) ={}&t(t-1)(t-\tilde{b})p^{\prime\prime}(t)~+ \\
	&\left( 2(t-1)(t-\tilde{b})+2t(t-\tilde{b})+t(t-1) \right)p^{\prime}(t).\notag
\end{align}
Recall that $\{t_m\}_{m=2}^{M-1}$ are also the roots of polynomial $p(t)$ of degree $M-2$, and therefore $r(t)$ must be divisible by $p(t)$, i.e., $r(t) = Q(t)p(t)$ for some polynomial $Q(t)$ of degree at most one. Hence, we have the following relationship between two polynomials as
\begin{equation}
	r(t) + \left( v_1t + v_0 \right) p(t) = 0,
	\label{eq:ode_undet_s}
\end{equation}
where $v_1$ and $v_0$ are the undetermined coefficients introduced to balance the degrees of polynomials.

Hence, the polynomial $p(t)$ satisfies the following second-order ODE as
\begin{equation}
	A(t)p^{\prime\prime}(t) + B(t)p^{\prime}(t) + V(t)p(t)=0,
	\label{eq:ODE_stieltjes}
\end{equation}
where 
\begin{equation}
	\begin{cases}
		A(t)\!\!\!\!&\!=t\left( t-1 \right)( t-\tilde{b} )\\
		B(t)\!\!\!\!&\!= 2\left( t-1 \right)( t-\tilde{b} ) + 2t(t-\tilde{b}) + t \left( t-1 \right)\\
		V(t)\!\!\!\!&\!= v_1t + v_0
	\end{cases}.
\end{equation}

In fact,~\eqref{eq:ODE_stieltjes} is known as the Heine-Stieltjes differential equation, and $V(t)$ is the corresponding Van Vleck polynomial~\cite{StieltjesPolynomials,stieltjes1898polynomial}. 
{Since the polynomial of degree $M$ on the left-hand side of~\eqref{eq:ODE_stieltjes} equals zero for all $t$, then all of the coefficients of the polynomial equals zero. Therefore, by comparing the coefficients of $t^{M-1}$ and $t^{M-2}$, we can solve the coefficients $v_1$ and $v_0$ 
of Van Vleck polynomial $V(t)$ as}
\begin{equation}
	\begin{cases}
		v_1=\!\!\!&-(M-2)(M+2)\\
		v_0=\!\!\!&M(M-2)+\tilde{b}(M-2)(M+1)\\
		&+(2M-1)c_{M-3}
	\end{cases},
	\label{eq:coeffs}
\end{equation}
where $c_{M-3}$ is an undetermined coefficient defined in~\eqref{eq:defp}. %

Note that the ODE in~\eqref{eq:ODE_stieltjes} can be reformulated as~\eqref{eq:heun}\addtocounter{equation}{1}, which corresponds to the canonical form of Heun's differential equation~\cite{Heun,StieltjesPolynomials} as
\begin{equation}
	p^{\prime\prime}\!(t)\!+\!\left( \frac{\gamma}{t}\!+\! \frac{\delta}{t\!-\!1}\!+\!\frac{\epsilon}{t\!-\!a_{\rm H}} \right)\!p^{\prime}\!(t) \!+\! \frac{\alpha\beta t-q}{t(t\!-\!1)(t\!-\!a_{\rm H})}p(t)\!=\!0,\notag
\end{equation}
where the standard Heun parameters are $\gamma = 2$, $\delta = 2$, $\epsilon = 1$, $a_{\rm H} = \tilde{b}$, $v_1 = \alpha\beta$, and $q = -v_0$, respectively. These parameters excluding $q$ are linked by the following Fuchsian relationship~\cite{Heun} as
\begin{equation}
	\gamma+\delta+\epsilon = \alpha+\beta+1,
	\label{eq:heunrelation}
\end{equation}
with which $\alpha$ and $\beta$ are solved interchangeably as
\begin{equation}
	\{ \alpha,\beta \} = \left\{ -(M-2),(M+2) \right\}.\label{eq:alphabeta} %
\end{equation}
{Up to this point, the original antenna position optimization problem ${\mathcal{P}_2}$ has been transformed into the problem of finding the roots of the polynomial $p(t)$ that satisfies the Heun's equation~\eqref{eq:heun}. In the following, our objective is to determine the coefficient $v_0$ of the Van Vleck polynomial $V(t)$ based on the properties of the Heun's equation and the physical constraints of the massive movable antenna placement problem. We then derive the corresponding polynomial solution and its roots, from which the optimal antenna positions can be obtained via the transformation relationship illustrated in Fig.~\ref{fig:eqeq}.}

\subsection{EVD-Based Efficient Antenna Position Acquisition}
\label{sec:EVD_sol}
In this subsection, we first derive the required conditions to obtain a desired truncated polynomial solution of Heun's equation~\eqref{eq:heun}. We then propose an efficient algorithm to obtain the desired polynomial solution, whose roots correspond to the equilibrium positions of antenna elements in the angular domain.

A given Heun's equation generally admits infinite solutions that can only be represented by the sum of infinite power series. To obtain a truncated polynomial solution, the conditions are derived in Proposition~\ref{prop:alphabetaq}.

\begin{prop}
	\label{prop:alphabetaq}
	Heun's equation admits a polynomial solution of degree $N$ if and only if
	\begin{enumerate}
		\item One of the parameters $\alpha$ and $\beta$ equals $-N$ for some non-negative integer $N$;
		\item The parameter $q$ is one of the eigenvalues of the tridiagonal matrix ${\bf R}\in\mathbb{R}^{(N+1)\times (N+1)}$, where
		\begin{equation}
			\renewcommand{\arraystretch}{0.7}
			\mathbf{R} =
			\begin{bmatrix}
				B_0 & C_0 & 0 & \cdots & 0 \\[-4pt]
				A_1 & B_1 & C_1 & \ddots & \vdots \\[-3pt]
				0 & A_2 & B_2 & \ddots & 0 \\[-2pt]
				\vdots & \ddots & \ddots & \ddots & C_{N-1} \\[2pt]
				0 & \cdots & 0 & A_N & B_N
			\end{bmatrix}.
			\label{eq:R}
		\end{equation}
		and 
		\begin{equation}
			\begin{cases}
				A_n\!\!\!\!&=(n-1+\alpha)(n-1+\beta)\\
				B_n\!\!\!\!&=-n\left( (n-1+\gamma)(1+a_{\rm H}) + a_{\rm H}\delta + \epsilon \right)\\
				C_n\!\!\!\!&=(n+1)(n+\gamma)a_{\rm H}
			\end{cases},
			\label{eq:three_term1}
		\end{equation}
	\end{enumerate}
	The $n$-th eigenvector ${\bf c}_n = [c_0^{(n)},c_1^{(n)},\cdots,c_{N}^{(n)} ]\in\mathbb{R}^{N+1}$ of matrix ${\bf R}$ directly corresponds to the polynomial coefficients of the $n$-th feasible polynomial solution $p_n(t)$.
\end{prop}
\begin{IEEEproof}
	See Appendix~\ref{append:3}.
\end{IEEEproof}

Applying Proposition~\ref{prop:alphabetaq} to our case with $N=M-2$, we obtain $M-1$ distinct eigenvalues $\{ q_n \}_{n=1}^{M-1}$, each associated with a coefficient vector ${\bf c}_n$ that yields a feasible polynomial solution $p_n(t)$. The roots of $p_n(t)$ can then be 
efficiently computed via EVD of the matrix
\begin{equation}
	\renewcommand{\arraystretch}{0.7}
	{\bf C} = \left[ \begin{matrix}
		0& 0 & \cdots&0 &-\tilde{c}_0^{(n)} \\[-3pt]
		1&0&\cdots&0 &-\tilde{c}_1^{(n)}\\[-2pt]
		0&1&\cdots&0&-\tilde{c}_2^{(n)}\\[-1pt]
		\vdots&\vdots&\ddots&\vdots&\vdots\\[-1pt]
		0&0&\cdots& 1 & -\tilde{c}_{M-3}^{(n)}
	\end{matrix}\right],
	\label{eq:C}
\end{equation}
where $\tilde{c}_i^{(n)} = c_i^{(n)}/c_{M-2}^{(n)}$. %
Note that the $M-1$ distinct polynomial solutions $\{ p_n (t) \}_{n=1}^{M-1}$ correspond to $M-1$ possible distributions of the roots. 

Recall in Fig.~\ref{fig:eqeq}(c) that the roots of the desired polynomial solution $p(t)$ must all lie within the interval $(0,1)$, it is therefore necessary to identify the specific one among all $M-1$ possible solutions.

In the following, we first introduce the Sturm-Liouville theory, which establishes the relationship between the eigenvalue $q$ of matrix ${\bf R}$ and the root distribution of the polynomial solutions to 
second-order ODEs such as Heun's equation.

\begin{theo}[Sturm-Liouville Theory~\cite{Sturm,zwillinger1998handbook}]\label{theo:SL}
	Consider the second-order linear ODE in the Sturm-Liouville form as
	\begin{equation}
		\frac{\rm d}{{\rm d} x}\left( f_1(x) \frac{{\rm d}}{{\rm d} x} y(x)\right) + f_2(x)y(x) + \lambda f_3(x)y(x) = 0,
		\label{eq:SLform}
	\end{equation}
	where $f_1(x)$, $f_2(x)$, and $f_3(x)$ are continuous functions on the finite regular interval $[x_{\rm L},x_{\rm R}]$, with $f_1(x) > 0$ and $f_3(x) > 0$ for all $x \in [x_{\rm L},x_{\rm R}]$. 
	A real-valued $\lambda$ that leads to non-trivial solutions\footnote{In the scope of Sturm-Liouville theory, the trivial solution refers to $y\equiv 0$.} to the problem~\eqref{eq:SLform} is called an eigenvalue of the Sturm-Liouville problem, while the corresponding solution $y(x)$ is called an eigenfunction.
	
	For any regular Sturm-Liouville problem where real eigenvalues can be ordered as
	\begin{equation}
		\lambda_1<\lambda_2<\cdots<\lambda_n<\cdots,
		\label{eq:eigenlist}
	\end{equation}
	unique eigenfunction $y_n(x)$ corresponding to the $n$-th eigenvalue has exactly $n-1$ zeros within the interval $[x_{\rm L},x_{\rm R}]$.
\end{theo}

\begin{figure}[t]
	\centering
	\vspace{-3mm}
	\includegraphics[width=0.425\textwidth]{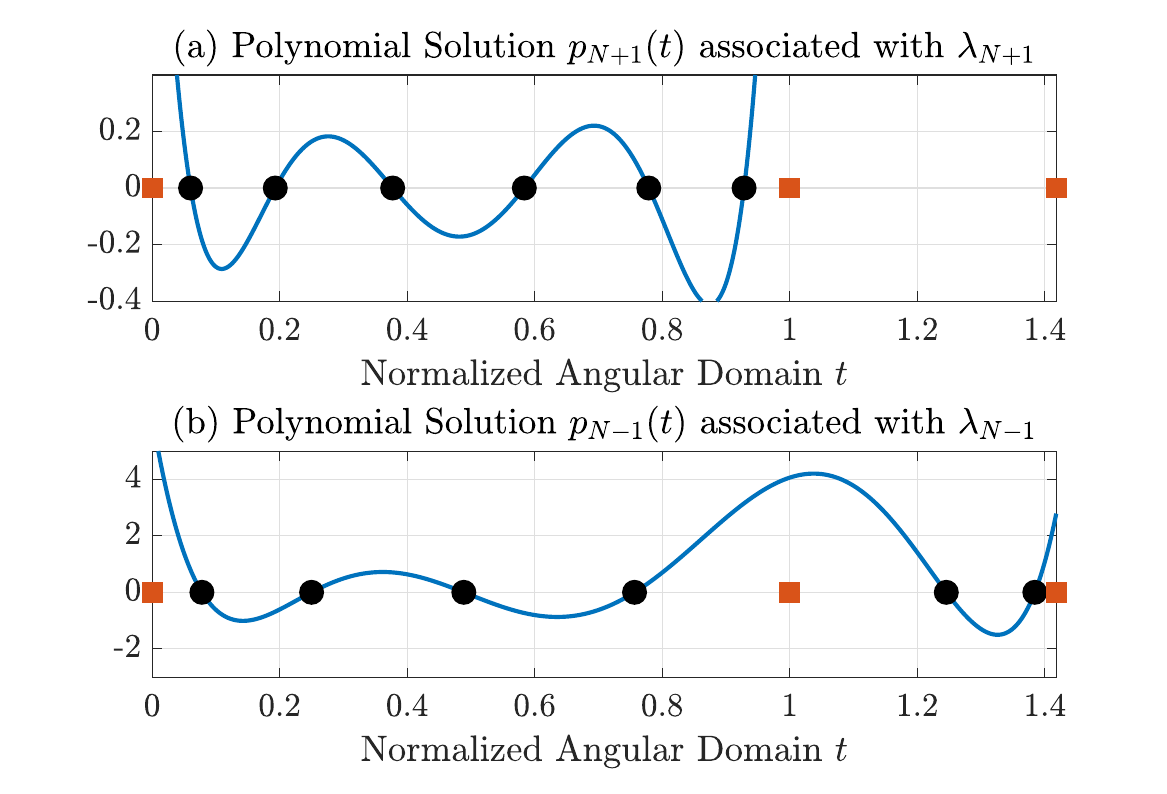}
	\vspace{-3mm}
	\caption{An illustration of polynomial solution $p_n(t)$ of Heun's equation~\eqref{eq:heun} with $a_{\rm H} = \tilde{b} \approx 1.41$ and $N=6$, associated with (a) the $7$-th eigenvalue (i.e., $\lambda_{N+1}$) and (b) the $5$-th eigenvalue (i.e., $\lambda_{N-1}$), as numbered in ascending order in~\eqref{eq:eigenlist}.}
	\label{fig:phys_imag}
	\vspace{-3mm}
\end{figure}

Theorem~\ref{theo:SL} reveals that, if the original Heun's equation admits the Sturm-Liouville form, then the number of roots of the polynomial solution $p(t)$ within the desired interval $(0,1)$ in the normalized angular domain, is directly connected to the \textit{order} of the corresponding eigenvalue, as illustrated in Fig.~\ref{fig:phys_imag}.

According to the property of the Heun's equation defined on the real-axis~\cite{StieltjesPolynomials}, all roots of the polynomial solutions $p(t)$ lie within the interval $[\min(0,\tilde{b}), \max(1,\tilde{b})]$. Therefore, based on the value of $\tilde{b}$, we categorize the problem into the following two cases.

\subsubsection{\bf Case with $\tilde{b}<0$ or $\tilde{b}>1$}

\begin{figure}[t]
	\centering
	\vspace{-3mm}
	\includegraphics[width=0.45\textwidth]{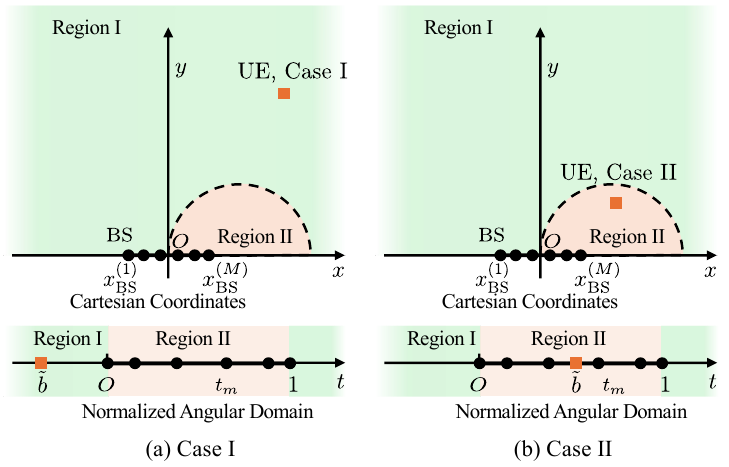}
	\vspace{-3mm}
	\caption{Cases of UE positions in the near-field region: Case I: The position of UE falls in region I, where $\tilde{b}<0$ or $\tilde{b}>1$ holds in the normalized angular domain; Case II: The position of UE falls in Region I, where $0<\tilde{b}<1$ holds in the normalized angular domain.}
	\label{fig:regions1}
	\vspace{-3mm}
\end{figure}

In this case, as illustrated in Fig.~\ref{fig:regions1}, our target is to find the proper eigenvalue that ensures all of the roots of the corresponding polynomial solution $p(t)$ lie within $(0,1)$. Given Theorem~\ref{theo:SL}, we can directly derive the following corollary.

\begin{corollary}\label{coro:eigen_sel1}
	For $\tilde{b}<0$ and $\tilde{b}>1$ corresponding to the region I in Fig.~\ref{fig:regions1}, the optimal choices of $q$ are respectively the \uline{largest} and \uline{smallest} eigenvalues of $\mathbf{R}$ in~\eqref{eq:R}.
\end{corollary}
\begin{IEEEproof}
	For $\tilde{b}<0$, the Heun's equation in~\eqref{eq:heun} can be reformulated into Sturm-Liouville form as~\eqref{eq:SLform} with
	\begin{equation}
		\begin{cases}
			f_1(t) \!\!\!\!&= t^2(t-1)^2(t-\tilde{b})\\[-1pt]
			f_2(t) \!\!\!\!&= t^2(t-1)v_1\\[-1pt]
			f_3(t) \!\!\!\!&= -t(t-1)\\[-1pt]
			\lambda_n \!\!\!\!& = q_n%
		\end{cases}.\label{eq:case1}
	\end{equation}
	Since we have $f_1(t)>0$ and $f_3(t)>0$ over $(0,1)$, the regular interval is $(0,1)$ according to Theorem~\ref{theo:SL}. To obtain $M-2$ zeros within the interval, we should select the $(M-1)$-th eigenvalue $\lambda_{M-1}$, i.e., the largest eigenvalue. Since we have $\lambda_n=q_n$ in~\eqref{eq:case1}, the optimal polynomial solution is the one associated with 
	the \uline{\it largest} eigenvalue $q_{M-1}$ of matrix ${\bf R}$ in~\eqref{eq:R}.
	
	For $\tilde{b}>1$, the Sturm-Liouville form of Heun's equation in~\eqref{eq:case1} is reformulated as
	\begin{equation}
		\begin{cases}
			f_1(t) \!\!\!\!&= -t^2(t-1)^2(t-\tilde{b})\\[-1pt]
			f_2(t) \!\!\!\!&= -t^2(t-1)v_1\\[-1pt]
			f_3(t) \!\!\!\!&= -t(t-1)\\[-1pt]
			\lambda_n \!\!\!\!& = -q_n%
		\end{cases}.\label{eq:case2}
	\end{equation}
	In this case, in order to select the largest eigenvalue $\lambda_{M-1}$, we have to choose the \uline{\it smallest} eigenvalue $q_1$ of matrix ${\bf R}$ in~\eqref{eq:R} due to $\lambda_n = -q_n$ in~\eqref{eq:case2}.
\end{IEEEproof}

Let the optimal choice of eigenvalue be $q^\star$, the desired polynomial coefficients are the elements of the corresponding eigenvector ${\bf c}^\star$. The roots $\{ t_m^\star \}_{m=2}^{M-1}$ of the polynomial solution associated with the coefficient vector ${\bf c}^\star$ can then be efficiently obtained by an EVD operation of the corresponding matrix ${\bf C}$ in~\eqref{eq:C}. 
The positions of antenna elements $\{ x_m^{\star} \}_{m=2}^{M-1}$ are then obtained by a denormalization of~\eqref{eq:tdomain} and an inverse coordinate transform of~\eqref{eq:sdomain}, as shown in Fig.~\ref{fig:eqeq}.

\subsubsection{\bf Case with $0<\tilde{b}<1$}

In this case, as illustrated in Fig.~\ref{fig:regions1}, for $M-1$ different eigenvalues $\{ q_n \}_{n=1}^{M-1}$ of $\bf R$, all the roots of the corresponding polynomial solution must lie within the interval $(0,1)$. 

Therefore, to find the optimal antenna positions that maximize the objective $J({\bf s})$ in $\mathcal{P}_2$, we first calculate all $M-1$ eigenvectors $\{{\bf c}_n \}_{n=1}^{M-1}$ that correspond to $M-1$ polynomial solutions, and then compute the corresponding roots $\{ t_m^{(n)} \}_{m=2}^{M-1}$ ($1\leq n\leq M-1$) of the polynomial solutions. We then denormalize the roots to the angular domain as $\{ s_m^{(n)} \}_{m=2}^{M-1}$, and select the one that satisfies
\begin{equation}
	\{ s_m^{(n^\star)} \}_{m=2}^{M-1} = \arg\underset{1\leq n\leq M-1}{\max } J\left( {\bf s}_n \right),\label{eq:eigensel}
\end{equation}
where ${\bf s}_n = [s_{\min},s_2^{(n)},\cdots,s_{M-1}^{(n)},s_{\max}]^T$.%
\vspace{3mm}

The procedure of calculating the antenna positions is summarized in \textbf{Algorithm~\ref{alg:opt}}.
\begin{remark} 
	For a wireless system equipped with $M = 32$ BS antennas, $N = 8$ UE antennas, and operating at carrier frequency $f_{c} = 10$~GHz, the area of Region~II, as illustrated in Fig.~\ref{fig:regions1}, is no greater than $0.07$~m$^2$. This implies that, in practical deployments, UEs are unlikely to lie within this region, and thus the potential computational overhead introduced by~\eqref{eq:eigensel} is typically negligible.
\end{remark}

\begin{algorithm}[t]
	\caption{Optimal Antenna Placement}\label{alg:opt}
	\begin{algorithmic}[1]
		\REQUIRE The number of antenna elements $M$, the aperture of BS array $A$, the centroid position $(x_0,y_0)$ and azimuth angle $\theta_{\rm UE}$ of the UE array.
		\ENSURE The positions $\{ x_{\rm BS}^{(m)} \}_{m=1}^{M}$ of $M$ antenna elements.
		\STATE Determine the parameters $s_1 = s_{\rm min}$, $s_{M} = s_{\rm max}$ according to~\eqref{eq:sdomain}, and $\tilde{b}$ according to~\eqref{eq:bpprime}.
		\STATE Determine the Heun parameters $\gamma = 2$, $\delta = 2$, $\epsilon = 1$, $a_{\rm H} = \tilde{b}$, $\alpha = -(M-2)$, $\beta = M+2$, and $v_1 = \alpha\beta$.
		\STATE Build matrix $\bf R$ according to~\eqref{eq:R} and~\eqref{eq:three_term1}.
		\STATE Perform EVD on ${\bf R}$ and sort the eigenvalues as
		\vspace{-1mm}
		\begin{equation}
			q_1<q_2<\cdots<q_{M-1}.\notag\vspace{-1mm}
		\end{equation} 
		\IF{$\tilde{b}<0$}
		\STATE Select the $(M-1)$-th eigenvector ${\bf c}_{M-1}$.
		\ELSIF{$\tilde{b}>1$}
		\STATE Select the $1$-st eigenvector ${\bf c}_{1}$.
		\ELSIF{$0<\tilde{b}<1$}
		\STATE Select the $n^\star$-th eigenvector ${\bf c}_{n}^\star$ according to~\eqref{eq:eigensel}.
		\ENDIF
		\STATE Build matrix ${\bf C}$ according to~\eqref{eq:C} with the selected eigenvector.
		\STATE Perform EVD of ${\bf C}$ and obtain the eigenvalues $\{ t_m \}_{m=2}^{M-1}$.
		\STATE Denormalize $\{ t_m \}_{m=2}^{M-1}$ to the angular domain by~\eqref{eq:tdomain} as
		\vspace{-1mm}
		\begin{equation}
			s_m = t_m\left( s_{\rm max}-s_{\rm min} \right) + s_{\rm min}.\notag\vspace{-1mm}
		\end{equation}
		\STATE Transform angular domain positions 
		\vspace{-1mm}
		\begin{equation}
			\{ s_{\rm min},s_{\rm max} \} \cup \{ s_m \}_{m=2}^{M-1}\notag\vspace{-1mm}
		\end{equation}
		to the Cartesian coordinates on $x$-axis by
		\vspace{-1mm}
		\begin{equation}
			x_{\rm BS}^{(m)} = x_0 + y_0 \tan\left(  \arcsin\left( s_m + \theta_{\rm UE} \right) \right).
			\label{eq:stox}
			\vspace{-1mm}
		\end{equation}
		\RETURN The positions on $x$-axis $\{ x_{\rm BS}^{(m)} \}_{m=1}^{M}$.
	\end{algorithmic}
\end{algorithm}

\section{Asymptotic Analysis}

In this section, we conduct asymptotic analysis for the massive MIMO system as $M \to \infty$ in the near field, and propose a simplified and computationally efficient {closed-form} antenna placement {solution}.

To begin with, as the UE locates in the near‑field region with centroid distance $r_{\rm Fres}\leq r_0\leq r_{\rm Ray}$ larger than aperture $A_{\rm T}$, where $r_{\rm Fres} = 0.5 \sqrt{A_{\rm T}^3/\lambda}$~\cite{7942128} and $r_{\rm Ray} = {2A_{\rm T}^2}/{\lambda}$, the admissible values of $s_m$ are restricted to an increasingly narrow range around $s_0 = \sin \left(\varphi_0 - \theta_{\rm UE}\right)$. In this case, according to~\eqref{eq:taylor1}, the external influence term in~\eqref{eq:js} gradually degenerates to a constant around
\begin{equation}
	\sum_{m=1}^{M}\log_2 \left( 1-\tilde{s}_m^2 \right) \simeq M\log\cos^2\varphi_0.
\end{equation}
Then, when the number of antennas $M\to\infty$, {the pairwise spacings $s_j - s_i$ become much smaller than $1-\tilde{s}_m^2$, rendering the external influence much less significant than the internal influence, i.e., }
\begin{equation}
	\sum_{1\leq i< j\leq M}\!\!\!\! \log_2 \left( s_j -s_i \right) \gg \sum_{m=1}^{M} \log_2 \left( 1-\tilde{s}_m^2 \right).
\end{equation}
Therefore, we can safely neglect the term associated with the external influences and approximate the equilibrium condition~\eqref{eq:equilibrium} by retaining only the dominant internal interaction term as
\begin{equation}
	\sum_{\substack{i=2\\i\neq m}}^{M-1}\frac{1}{s_m-s_i}  + \frac{1}{s_m-s_{\rm min}}+\frac{1}{s_m-s_{\rm max}}=0,
	\label{eq:equilibrium_jacobi}
\end{equation}
which further yields the normalized equilibrium equation
\begin{equation}
	\sum_{\substack{i=2\\i\neq m}}^{M-1}\frac{1}{t_m-t_i}  =- \frac{1}{t_m+1}-\frac{1}{t_m-1},
	\label{eq:equilibrium_jacobi_t}
\end{equation}
with the normalization operation
\begin{equation}
	t = \frac{2s-\left( s_{\rm min}+s_{\rm max} \right)}{\left( s_{\rm max}-s_{\rm min} \right)}\in(-1,1).
\end{equation}
Similar to the derivations in~\eqref{eq:equilibrium_2}--\eqref{eq:ODE_stieltjes},~\eqref{eq:equilibrium_jacobi_t} gives rise to a differential equation as
\begin{equation}
	\left(1-t^2\right)p^{\prime\prime}(t) - 4tp^{\prime}(t) + v_0 p(t) = 0,
	\label{eq:jacobiODE}
\end{equation}
where $v_0 = (M-2)(M+1)$ by comparing the coefficients of $t$ with the highest order. It is worth noting that ~\eqref{eq:jacobiODE} is known as the \textit{Jacobi differential equation}~\cite{zwillinger1998handbook,szego1939orthogonal}
\begin{equation}
	\left(1-t^2\right)p^{\prime\prime}(t) +\left( \beta-\alpha-\left(\alpha + \beta+2 \right)t \right) p^{\prime}(t) + \lambda p(t) = 0\notag
\end{equation}
with $\lambda=(M-2)(M-1+\alpha+\beta)$, and therefore the corresponding polynomial solution is the \textit{Jacobi polynomial} $P_{M-2}^{(\alpha,\beta)}(t)$ of degree $M-2$. In our case, we have $\alpha=\beta = 1$, and the corresponding polynomial solution is then $P_{M-2}^{(1,1)}(t)$.

\begin{prop}\label{prop:cheby}
	As $N\to \infty$, the zeros of Jacobi polynomials $P_{N}^{(1,1)}(x)$ with $\alpha = \beta = 1$ converge to a fixed solution as
	\begin{equation}
		x_n = \cos\left( \frac{\left( 2n+1 \right)\pi}{2N} \right), \quad n=0,1,\cdots,N-1.
		\label{eq:zeroscheby}
	\end{equation}
\end{prop}
\begin{IEEEproof}
	See Appendix~\ref{append:4}.
\end{IEEEproof}

With Proposition~\ref{prop:cheby}, the asymptotically optimal antenna positions $\{ t_m \}_{m=2}^{M-1}$ on the normalized angular domain as $M\to \infty$ is defined by the zeros of $P_{M-2}^{(-\frac{1}{2},-\frac{1}{2})}(t)$, which can be efficiently obtained according to~\eqref{eq:zeroscheby} as $t_{m+1} = x_m$ for $m = 1,\cdots, M-2$ together with $t_M = -t_1 = 1$. The positions in the angular domain are then transformed as
\begin{equation}
	s_m = \frac{\left( s_{\rm max}-s_{\rm min} \right)}{2} t_m + \frac{\left( s_{\rm max}+s_{\rm min} \right)}{2},
	\label{eq:ttos}
\end{equation}
and are further transformed to the coordinates by~\eqref{eq:stox}. 
The detailed steps are summarized in \textbf{Algorithm~\ref{proc:1}}.
\begin{algorithm}[t]
	\caption{Asymptotically Optimal Positions at $M\to\infty$}\label{proc:1}
	\begin{algorithmic}[1]
		\STATE Calculate the roots of the Jacobi polynomial $P_{M-2}^{(-\frac{1}{2},-\frac{1}{2})}(x)$ according to~\eqref{eq:zeroscheby} as $\{ t_m \}_{m=2}^{M-1}$.
		\STATE Complete the solution by 
		\vspace{-1mm}
		\begin{equation}
			\{ t_m \}_{m=1}^{M}\!=\! \{ t_m \}_{m=2}^{M-1} \cup \{ -1,1 \}.\notag
			\vspace{-1mm}
		\end{equation}
		\STATE Transform to the angular domain according to~\eqref{eq:ttos}.
		\STATE Transform angular domain solutions $\{ s_m \}_{m=1}^{M}$ to the coordinates on $x$-axis by~\eqref{eq:stox}.
		\RETURN Optimal antenna positions $\{ x_{\rm BS}^{(m)} \}_{m=1}^{M}$.
	\end{algorithmic}
\end{algorithm}

\begin{remark}
	Unlike the solutions in Algorithm~\ref{alg:opt}, the roots of Jacobi polynomials of a given degree $M$ can be computed and cached in advance, thereby avoiding repeated EVD operations and resulting in a much lower computational complexity.
\end{remark}

\section{Numerical Results}

\subsection{Simulation Setup}
\label{sec:setup}
{In the} simulation, the carrier frequency is set as $f_{\rm c} = 10$ GHz. {The Fresnel boundary is defined by $r_{\rm Fres} = 0.5\sqrt{A_{\rm T}^3/{\lambda}}$,} and the Rayleigh distance is defined by $r_{\rm Ray} = {2A_{\rm T}^2}/{\lambda}$, where $A_{\rm T} = (M-1)d$ denotes the moving range of the antenna elements with $d = 2\lambda$. 
The UE is equipped with a ULA with $N=8$ fixed-position antennas, and the standard antenna spacing is $\lambda/2$. The centroid of the UE array is selected randomly from the near-field region within centroid distance $r_0 \in [r_{\rm Fres}, r_{\rm Ray}]$ and azimuth angle $\varphi_0 \in [-\pi/3, \pi/3]$, respectively. The azimuth angle of the UE array is also randomly selected within $\theta_{\rm UE} \in [-\pi/3, \pi/3]$, and we ensure that $\varphi_m - \theta_{\rm UE}\in\left[ -\pi/2,\pi/2 \right]$ for all $1\leq m\leq M$. 
As the received power is significantly affected by the distance, especially in the near-field region, we normalize the SNR to $\rho = 20$ dB. 

{For comparison purposes, we consider the following} 
antenna configuring schemes {at BS}.
\begin{itemize} 
	\item {\bf Random}: In each simulation step, $1,000$ antenna placement patterns are randomly generated at the BS for SE evaluation.
	\item {\bf Greedy Selection}~\cite{10458417}: The antenna positions are successively selected from ${G}_{\rm AS}=2 M$ uniform grids by orthogonal matching pursuit.
	\item {\bf Gradient Descent} (GD)~\cite{10354003}: The antenna positions are obtained by solving the original convex problem $\mathcal{P}_2$ by the numerical gradient descent method.
	\item {\bf Proposed EVD}: The antenna positions are obtained by employing \textbf{Algorithm~\ref{alg:opt}}. %
	\item {\bf Proposed Closed-form}: The antenna positions are obtained using the closed-form solution in \textbf{Algorithm~\ref{proc:1}}.
\end{itemize}
\subsection{Approximation Accuracy}
\begin{figure}[t]
	\vspace{-3mm}
	\centering
	\includegraphics[width=0.425\textwidth]{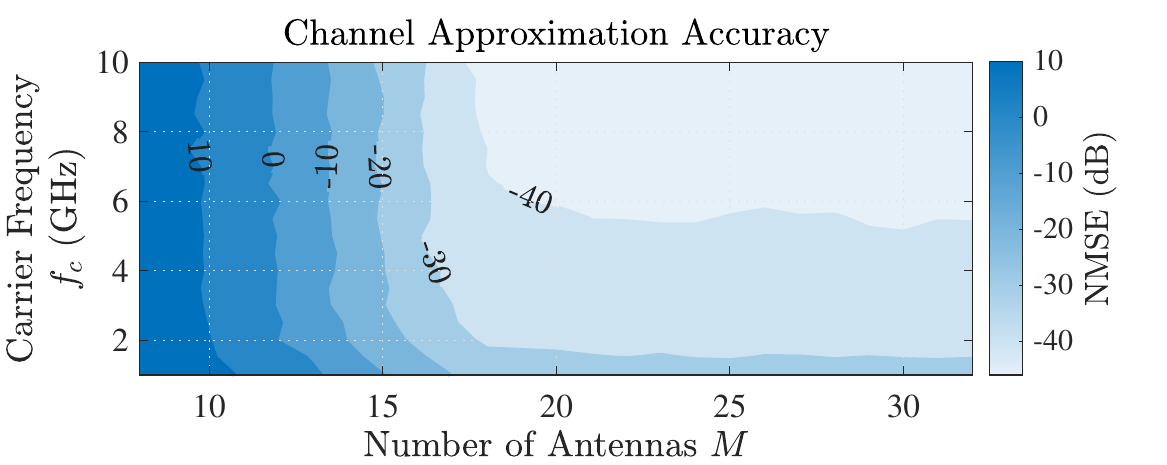}
	\vspace{-1mm}
	\caption{The approximation error between ${\bf H}_{\bf x}$ in~\eqref{eq:chmodel_los} and the truncated approximation $\uline{\bf H}_{\bf x} = \uline{\bf P}_M {\bf D}_{\rm T}$ in~\eqref{eq:truncate_}.}
	\label{fig:approx_acc}
	\vspace{-3mm}
\end{figure}

We first validate the accuracy of the approximation adopted in the problem formulation in Section~\ref{sec:formulation}, where the infinite-column matrices are truncated to $M$ columns. Since ${\bf V}_{\rm R}$ also depends on carrier frequency $f_{\rm c}$, we evaluate the approximation error with respect to both $M$ and $f_{\rm c}$ in contour manner. Under the setup specified in Section~\ref{sec:setup}, we perform $2,000$ Monte Carlo trials and compute the average normalized mean squared error (NMSE) as
\begin{equation}
	{\rm NMSE}\left( {\bf H}_{\bf x}, \uline{\bf H}_{\bf x} \right) = \frac{\left\Vert {\bf H}_{\bf x}- \uline{\bf H}_{\bf x} \right\Vert_{F}^2}{\left\Vert {\bf H}_{\bf x} \right\Vert_{F}^2},
\end{equation}
where $\uline{\bf H}_{\bf x} = \uline{\bf P}_M {\bf D}_{\rm T}$ denotes the truncated approximation of channel matrix ${\bf H}_{\bf x}$, and $\Vert\cdot\Vert_F$ represents the Frobenius norm. 
As illustrated in Fig.~\ref{fig:approx_acc}, the approximation error decreases rapidly as $f_{\rm c}$ and $M$ increase, which demonstrates the accuracy of the adopted approximation in massive MIMO systems. Therefore, unless otherwise stated, we consider $M\ge 16$ antenna elements in the subsequent simulations, where the expected approximation NMSE remains lower than $-20$ dB.

\subsection{Spectral Efficiency Performance}

\begin{figure}[t]
	\vspace{-3mm}
	\centering
	\includegraphics[width=0.425\textwidth]{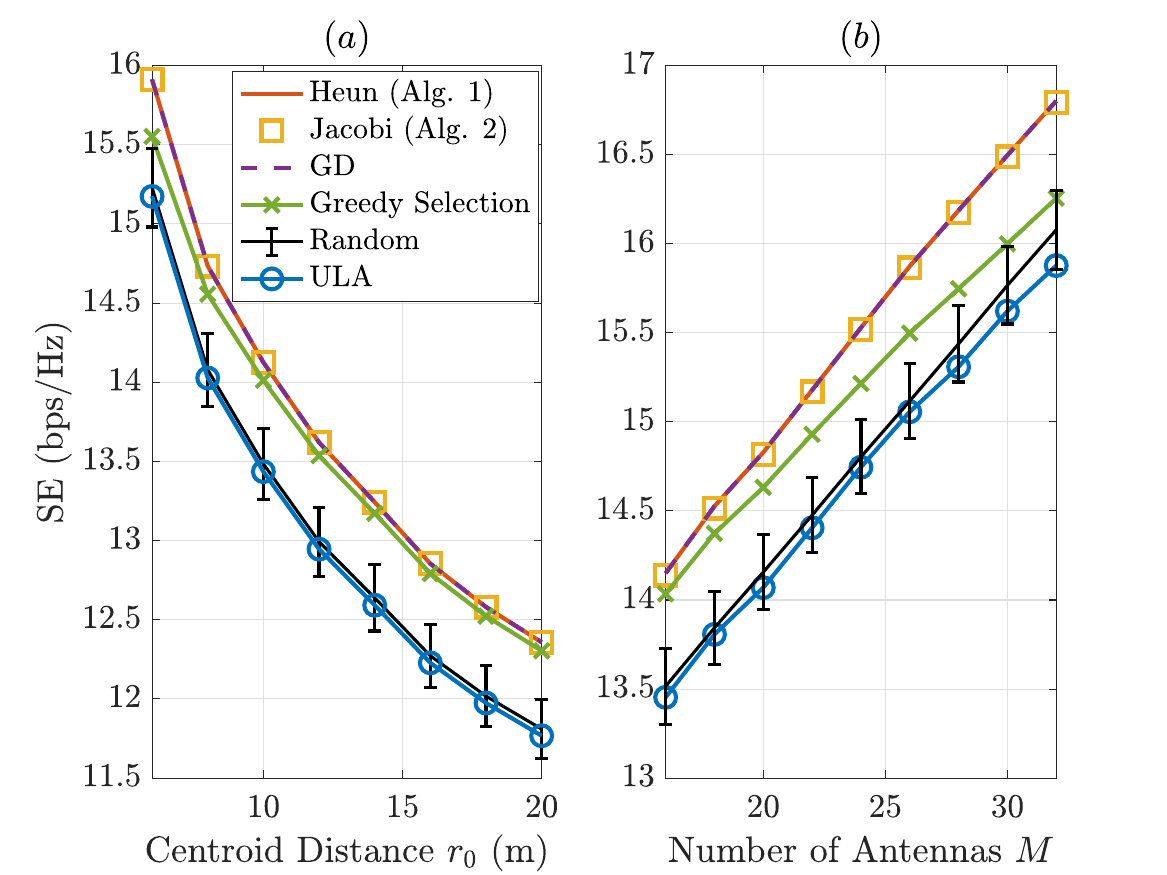}
	\caption{The illustration of SE performance with (a) varying centroid distance $r_0\in [5,20]$ m when $M = 16$ and (b) varying number of antennas $M$ when $r_0 = 10$~m.}
	\label{fig:SE}
	\vspace{-3mm}
\end{figure}

We then investigate the SE performance of the proposed antenna placement schemes with respect to the near-field distance $r_0$ and number of antennas $M$. As shown in Fig.~\ref{fig:SE}(a), with $M=16$ movable antenna elements, the proposed antenna placement schemes (i.e., Algorithm~\ref{alg:opt} and Algorithm~\ref{proc:1}), achieve the highest SE performance within the near-field region $r_0\in[5,20]$ m, which coincides with the GD benchmark that gives the optimal solution. As $r_0$ increases, the performance advantage of the proposed schemes over the ULA baseline and the greedy selection scheme is well preserved. Note that the greedy selection scheme suffers from noticeable SE degradation at shorter distances, which stems from the inequivalence between the greedy local optimum and the global optimum.

Moreover, as shown in Fig.~\ref{fig:SE}(b), when centroid distance $r_0 = 10$ m, the SE achieved by all considered schemes increases with the number of antenna elements $M$. The proposed schemes again attain the highest SE performance and coincide with the GD benchmark, and their performance gain over other baseline schemes continues to increase with $M$. On the contrary, the greedy scheme exhibits an SE loss in the massive MIMO regime, which is mainly attributed to its on-grid searching mechanism that fails to capture the true optimal antenna configuration in large antenna arrays. Hence, the numerical results in Fig.~\ref{fig:SE} validate the performance advancements of the proposed EVD-based efficient algorithm, and the consistency of the proposed asymptotic closed-form solution.

\begin{figure}[t]
	\centering
	\vspace{-3mm}
	\includegraphics[width=0.425\textwidth]{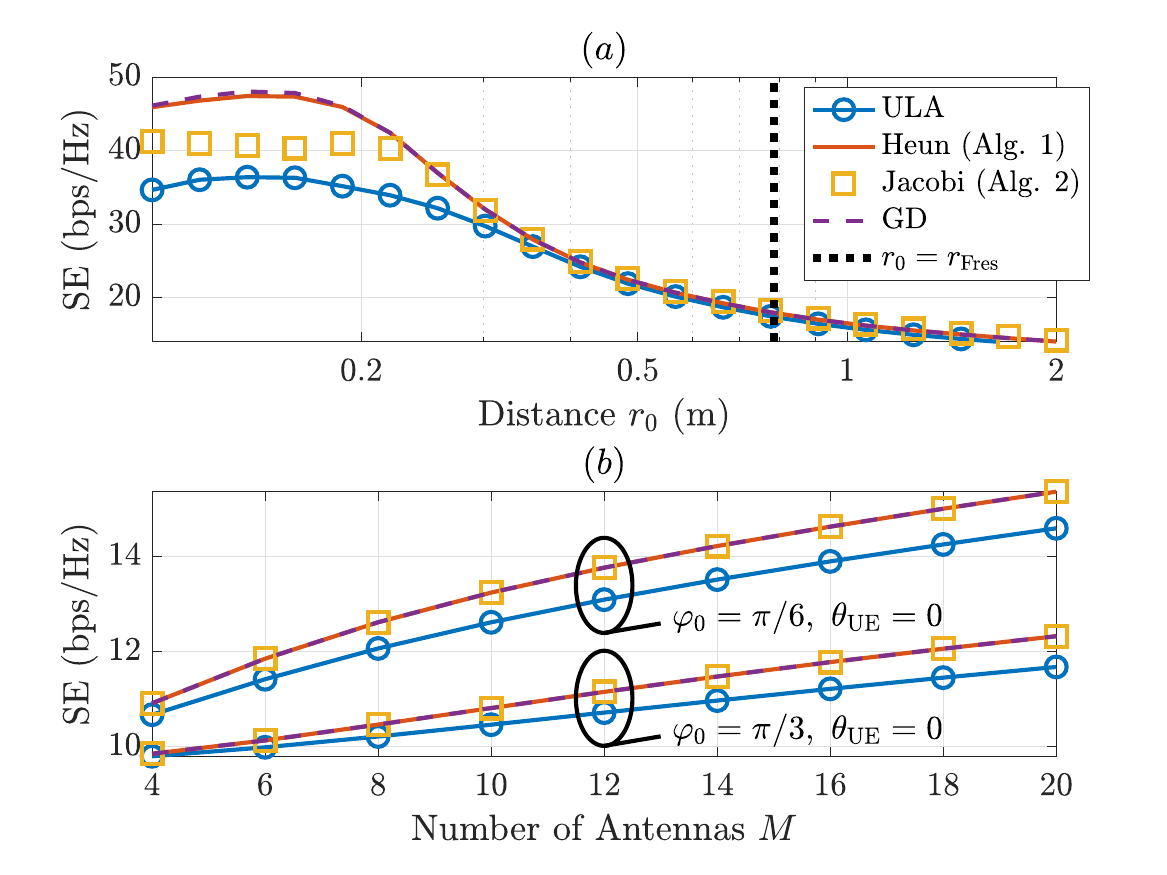}
	\vspace{-3mm}
	\caption{The illustration of the asymptotic behavior with respect to (a) centroid distance $r_0$ wgeb $\theta_{\rm UE} =0$ and $\varphi_0 = \pi/3$ and (b) the number of antennas $M$ when $\theta_{\rm UE} =0$ and $\varphi_0  \in \{ \pi/6, \pi/3\}$.}
	\label{fig:approx1}
\end{figure}

To further verify the asymptotic behavior of the proposed schemes, we conduct simulations under relatively small centroid distances $r_0\in[0.1,2]$ m and numbers of antennas $4\leq M\leq 16$. As illustrated in Fig.~\ref{fig:approx1}(a), when the distance $r_0$ is smaller than the Fresnel near-field boundary $r_{\rm Fres}=0.79$ m, the proposed asymptotic scheme, i.e., Algorithm~\ref{proc:1}, exhibits a slight performance loss compared to Algorithm~\ref{alg:opt}, which quickly diminishes as $r_0$ approaches the aperture {$A_{\rm T}$ of BS. In addition, we investigate the asymptotic behavior with respect to $M$. As depicted in Fig.~\ref{fig:approx1}(b), the performance of the proposed asymptotic scheme in Algorithm~\ref{proc:1}} 
also shows more favorable performance than ULA, which is nearly identical to the accurate solution {of Algorithm~\ref{alg:opt}} starting from $M=4$ for both the considered setups. These observations further confirm that, even though the asymptotic solutions are derived when $M\to\infty$, the obtained asymptotic closed-form design is effective in small antenna size regimes at a practical near-field distance.

\subsection{Antenna Position Patterns}

\begin{figure}[t]
	\centering
	\includegraphics[width=0.425\textwidth]{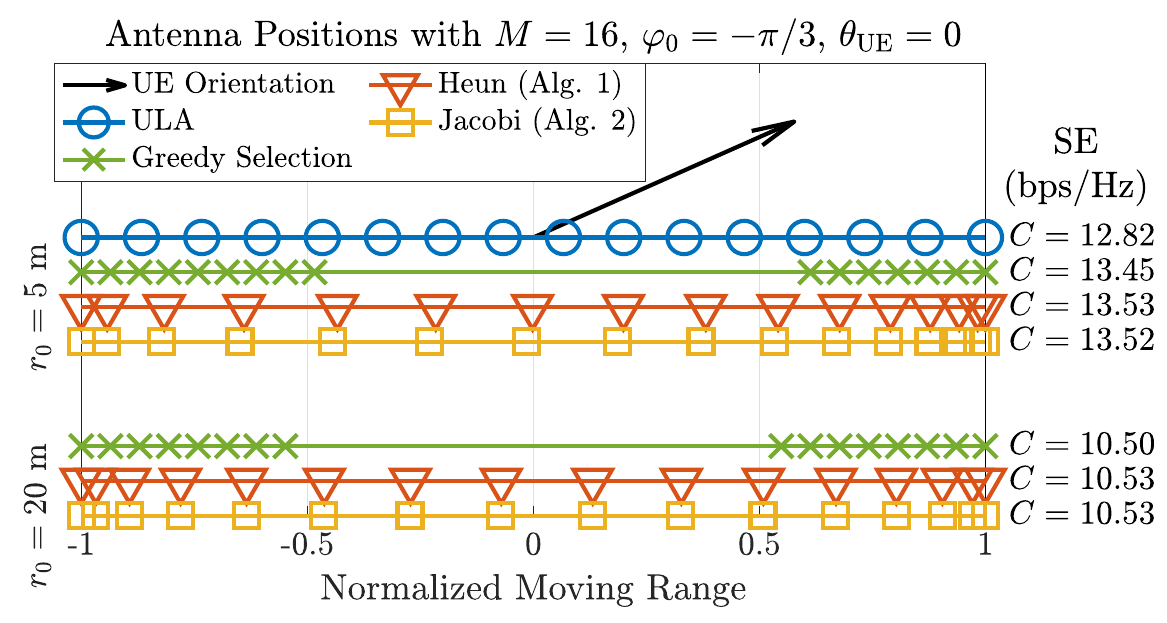}
	\vspace{-3mm}
	\caption{The antenna position patterns of different antenna placement schemes with $M=16$, $\varphi_0 = \pi/3$, and $\theta_{\rm UE}$ at centroid distance $r_0 \in \{ 5,20 \}$ m.}
	\label{fig:pos2}
	\vspace{-2mm}
\end{figure}

We also investigate the antenna position patterns for given specific system parameters. As shown in Fig.~\ref{fig:pos2}, the antenna pattern consistently exhibits a denser antenna placement near the array edges and a sparser placement around the center, while the greedy method fails to allocate any antenna in the middle. This observation suggests that the spatial propagation characteristics in the near-field region may vary more sharply at the array boundaries than at its center, thereby requiring denser spatial sampling at the edges to fully exploit the spatial DoF~\cite{liu2025nearfieldcommunicationmassivemovable}. In addition, the antenna elements tend to shift and cluster towards the UE-oriented side of the array, while preserving its overall dense-edge spatial structure. 
{This behavior can also be interpreted from Fig.~\ref{fig:eqeq}(a) with the corresponding first-order condition in~\eqref{eq:firstorder_ori}, where the external influence term exerts an attractive force that attracts the charges (i.e., the antennas) 
towards the UE direction, while its impact gradually diminishes as the link distance $r_0$ and the number of antennas $M$ increase.}

\subsection{Robustness to CSI Mismatch}
\begin{figure}[t]
	\centering
	\vspace{-3mm}
	\includegraphics[width=0.425\textwidth]{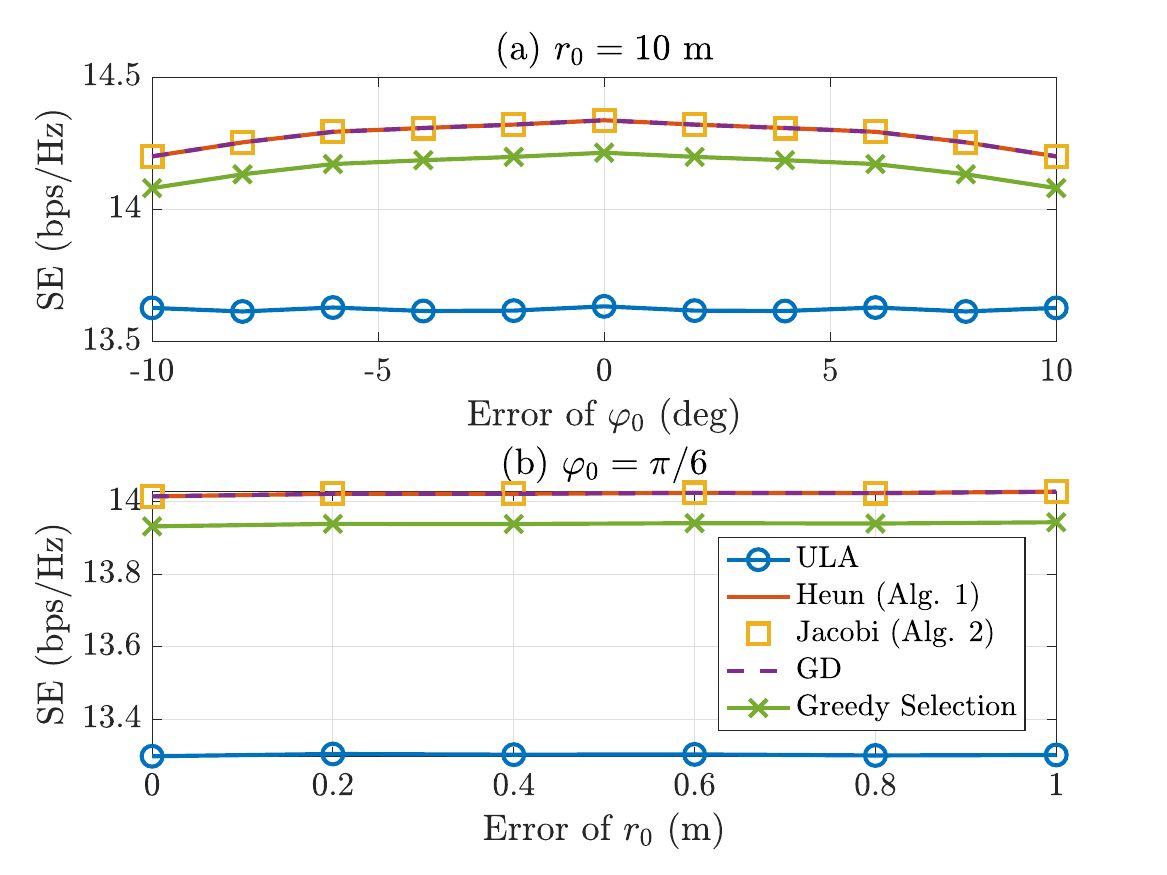}
	\vspace{-3mm}
	\caption{The robustness of different antenna placement schemes against estimation errors with respect to (a) UE azimuth angle $\varphi_0$ and (b) centroid distance $r_0$.}
	\label{fig:rob}
\end{figure}
We further evaluate the robustness of the proposed algorithms against the estimation error of the UE azimuth angle $\varphi_0$ and centroid distance $r_0$. {We first fix the centroid distance at $r_0 = 10$ m and test the impact of angle mismatch from $-10^\circ$ to $10^\circ$, {and the results are illustrated in Fig.~\ref{fig:rob}(a)}. 
As the absolute angle error increases, both the greedy selection and the proposed schemes exhibit performance degradation. However, within an error range of up to $10^\circ$ (around $60\lambda$ position error), the SE performance of the proposed schemes only drops $0.96\%$ %
and consistently remains the highest among all schemes.}
We then test the robustness to the distance mismatch, where the centroid distance ranges within $r_0 \in [10, 15]$ m while $\varphi_0$ is fixed at $\pi/6$. As shown in Fig.~\ref{fig:rob}(b), all schemes show strong robustness to distance errors, and the proposed algorithms consistently achieve the highest SE among all schemes. These observations confirm the robustness of the proposed design against inaccuracies in the prior distance and angle information.

\subsection{Computational Complexity}
\begin{table}[t]
	\begin{threeparttable}
	\centering
	\caption{Computational Complexity (Number of Multiplications) of different Methods.}
	\label{tab:complexity}

	\begin{tabular}{c|c|cc}
		\hline\hline
		\multirow{2}{*}{\bf Methods} & \multirow{2}{*}{\bf \begin{tabular}[c]{@{}c@{}}Computational \\ Complexity  ($\mathcal{O}$)\end{tabular}} & \multicolumn{2}{c}{\bf Avg. Execution Time (s)} \\ \cline{3-4} 
		&                                           &     $\boldsymbol{M=16}$         &      $\boldsymbol{M=64} $          \\ \hline
		{Algorithm~\ref{alg:opt}} &    $2M^3$\tnote{$\dagger$}                                       &        $7.39\!\times\!10^{-5}$         &      $9.07\!\times\!10^{-2}$          \\ \hline
		{Algorithm~\ref{proc:1}} &            N/A (Closed-form)                                &      \!\!\!\!$\bf 2.64\!\times\!10^{-7}$\!\!\!\!           &     \!\!\!$\bf 4.09\!\times\!10^{-7}$\!\!\!           \\ \hline
		\!\!GD~\cite{10354003}\!\!&     \!\!\!\!$\sim M^3$\!\!\!\!                                      &   $5.19\!\times\!10^{-2}$              &  \!\!$2.74$\!\!              \\ \hline
		Greedy~\cite{10458417}&     \!\!\!\!$2M(N\!+\!MN^2)\!+\!MN^3$\!\!\!\!                                      &   $1.09\!\times\!10^{-2}$              &  \!\!\!\!$2.24\!\times\!10^{-1}$\!\!\!\!              \\ \hline
		\hline
	\end{tabular}
	\begin{tablenotes}
		\item [$\dagger$] EVD can benefit from a variety of hardware-friendly algorithm implementations to reduce the execution time.
	\end{tablenotes}
	\vspace{-3mm}
\end{threeparttable}
\end{table}

We finally evaluate the computational complexity of the proposed algorithms and benchmarks~\cite{10663427,10458417}, along with the corresponding average execution time over $1,000$ test cases. The results are summarized in Table~\ref{tab:complexity}, where the numbers are rounded to three significant figures. Compared with the GD and greedy antenna selection schemes, {the} proposed Algorithm~\ref{alg:opt} shows a similar level of computational complexity, but the average execution time is significantly lower. This is because the EVD operations over structured matrices (e.g., the tri-diagonal form of $\bf R$ in~\eqref{eq:R}) {can be} highly optimized with hardware-friendly algorithms, while the gradient-based and greedy algorithms require repeated internal iterations. Moreover, with the asymptotic closed-form solution in Algorithm~\ref{proc:1}, the required execution time can be further reduced to the order of $10^{-7}$ s ($0.1~\mu{\text s}$), which is highly negligible. 
Overall, the proposed two-step EVD-based scheme offers a favorable trade-off between the computational burden and SE performance, making it particularly attractive for wireless systems with a relatively smaller number of antennas. In addition, the asymptotic closed-form solution in Algorithm~\ref{proc:1} attains microsecond-level execution time while preserving near-optimal performance, which renders it highly suitable for latency-critical and resource-constrained MA placement applications with an extremely large number of antennas.

\section{Conclusions}

In this paper, we investigated the optimal antenna placement problem and proposed a closed-form solution based on a two-step EVD operation, along with an asymptotic closed-form expression tailored for massive MIMO scenarios. Specifically, we reformulated the SE maximization problem into a convex weighted Fekete problem and revealed {that the optimal antenna configuration is fundamentally governed by the electrostatic equilibrium points}. 
To efficiently solve for the equilibrium points, we further reformulated the optimality conditions into Heun's equation and developed a two-step EVD algorithm based on Sturm-Liouville theory. %
The asymptotic behavior of Heun's equation was also analyzed, yielding a closed-form solution expressed in simple trigonometric functions. Simulation results demonstrated that the proposed schemes effectively exploit the spatial DoFs in near-field channels, {achieving substantial SE gains, exhibiting robustness to CSI mismatch}, while incurring negligible computational complexity and execution time.

\appendices
\section{Proof of Lemma~\ref{lemma:concave}}\label{append:1}
Let ${\bf J}$ be the Hessian matrix of the objective function $J({\bf s})$ in~\eqref{eq:js}, the $m$-th diagonal element of ${\bf J}$ is given by
\begin{align}
	&{\bf J}[m,m]= \frac{\partial^2 J}{\partial s_m^2}\label{eq:hessian_diag}\\ 
	={}&\!-\!\frac{1}{\ln 2}\!\left( \sum_{\substack{i=1\\i\neq m}}^{M} 
	\!\frac{1}{\left( s_i-s_m \right)^2}\!+\! \frac{2+\sin2\varphi_m\tan\left(\varphi_m\!-\!\theta_{\rm UE}\right)}{\left( 1-s_m^2 \right)\cos^2 \varphi_m}\! \right),\notag%
\end{align}
while the $(m,n)$-th off-diagonal element is given by
\begin{equation}
	{\bf J}[m,n] = \frac{\partial^2 J}{\partial s_m \partial s_n} =  \frac{1}{\left( s_m-s_n \right)^2 \ln 2}.
\end{equation}
The quadratic form of the Hessian matrix ${\bf J}$ given an arbitrary non-zero vector ${\bf v} = [v_1,v_2, \cdots,v_M]^T$ is then given by
\begin{align}
	& Q({\bf v})= {\bf v}^T {\bf J} {\bf v}\label{eq:quadform}\\ 
	={}&\!\! -\!\!\!\!\!\!\sum_{1\leq i<j\leq M}\!\! \frac{\left( v_i\!-\!v_j \right)^2}{\left( s_i \!-\! s_j \right)^2} \!-\! \sum_{m=1}^{M}\! \frac{(2\!+\!\sin2\varphi_m\tan\left( \varphi_m\!-\!\theta_{\rm UE} \right) )v_m^2}{\left( 1-s_m^2 \right)\cos^2 \varphi_m},\notag
\end{align}
where the factor $1/\ln 2$ is ignored for brevity. Let $u_m = \varphi_m - \theta_{\rm UE}$, the equality
\begin{align}
	&2+\tan u_m \sin2\varphi_m\\
	={}&\frac{\sec u_m}{2}\!\left(  \frac{4}{\sec u_m} \!-\!\cos\left( 2\varphi_m \!+\! u_m \right) \!+\!\cos\left( 2\varphi_m\!-\!u_m \right) \!\right)\notag
\end{align}
holds since we have $\varphi_m-\theta_{\rm UE}\in[-\pi/2,\pi/2]$ by definition. Hence, the quadratic form $Q({\bf v})$ in~\eqref{eq:quadform} is strictly negative for any given $\mathbf{v} \neq \mathbf{0}$.

Therefore, the Hessian matrix ${\bf J}$ is negative definite, i.e., ${\bf J}\prec 0$, which establishes that $J(\mathbf{s})$ is strictly concave. 
Moreover, as the feasible set defined by $s_1 < s_2 < \cdots < s_M$, where $s_1$ and $s_M$ fixed, is a convex set, $\mathcal{P}_2$ is hence a convex optimization problem.
\hfill\IEEEQEDhere

\section{Proof of Proposition~\ref{prop:alphabetaq}}\label{append:3}
Without loss of generality, suppose ${p}(t)$ is an infinite power series as
\begin{equation}
	{p}(t) = \prod_{n=1}^{\infty} \left( t - t_n \right) = \sum_{n=0}^{\infty} {c}_n t^n.
	\label{eq:newpoly2}
\end{equation}
where ${c}_n$ are the coefficients of $t^n$ for all $n\geq 0$. Substituting~\eqref{eq:newpoly2} into the Heun's equation~\eqref{eq:heun}, we obtain the following three-term recursion as
\begin{equation}
	A_n c_{n-1} + (B_n - q) c_n + C_n c_{n+1} = 0\quad (n\geq 1),
	\label{eq:three_term0}
\end{equation} 
where $A_n$, $B_n$, and $C_n$ are the recursion coefficients defined in~\eqref{eq:three_term1}, and we set $c_{-1} = 0$ according to the boundary convention~\cite{Heun}.

For the series to terminate at degree $N$ and yield a polynomial solution, it is necessary and sufficient that either $\alpha = -N$ or $\beta = -N$ for some non-negative integer $N$, which forces $A_{N+1} = 0$ and hence $c_r = 0$ for all $r > N$. This ensures the recurrence truncates at $r = N+1$.

Since we already have a non-positive integer solution for $\alpha$ from~\eqref{eq:heunrelation} as $\alpha = -N = -(M-2)$, the recursion loses $A_n$ at $A_{N+1} = (N+\alpha)(N+\beta) = 0$. However, for the series to terminate and yield a polynomial of degree $N$, it is additionally required that $c_n = 0$ for all $n>N$.

Let ${\bf c} = [c_0,c_1,\cdots,c_N]^T$ truncate at $r=N+1$, the recurrence form~\eqref{eq:three_term0} can be equivalently reformulated in matrix form as 
\begin{equation}
	\left({\bf R}-q{\bf I}_{N+1}\right){\bf c} = {\bf 0},
	\label{eq:eigform1}
\end{equation}
with ${\bf R}$ in the form of~\eqref{eq:R}. Note that~\eqref{eq:eigform1} takes the standard form of an eigenvalue equation. Hence, the truncated polynomial solution can be obtained if and only if the corresponding values of $q$ admit to the eigenvalues of matrix~${\bf R}$, and each corresponding eigenvector ${\bf c}$ defines the polynomial coefficients of a feasible polynomial solution $p(t)$.\hfill\IEEEQEDhere

\section{Proof of Proposition~\ref{prop:cheby}}\label{append:4}

{As a direct consequence of the three-term recursive relationship of orthogonal polynomials~\cite{szego1939orthogonal}, the roots of Jacobi polynomial $P_{N}^{(1,1)}(t)$ can be efficiently obtained by a single EVD operation of the Jacobi matrix
\begin{equation}
	\renewcommand{\arraystretch}{0.7}
	{\bf J}_{N}^{({1},{1})} = \left[
	\begin{matrix}
		0&d_1&0&\cdots&0\\[-4pt]
		d_1& 0&d_2&\ddots&\vdots\\[-3pt]
		0& d_2&0&\ddots&0\\[-2pt]
		\vdots&\ddots&\ddots&\ddots&d_{N-1}\\
		0&\cdots&0&d_{N-1}&0
	\end{matrix}\right],
	\label{eq:jacobimat}
\end{equation}
where $d_n = \sqrt{n(n+2)/\left( \left(2n+3\right)\left(2n+1\right) \right)}$, 
and the $N$ eigenvalues of the matrix ${\bf J}_{N}^{({1},{1})}$ are the corresponding roots of the Jacobi polynomial $P_{N}^{({1},{1})}(t)$. 

As $N \to \infty$, the off-diagonal elements $d_n$ in~\eqref{eq:jacobimat} converge to
\begin{equation}
	\lim_{n\to \infty} \sqrt{\frac{n\left(n+2\right)}{\left(2n+3\right)\left(2n+1\right)}} = \frac{1}{2}.
\end{equation}
Consequently, the matrix ${\bf J}_{N}^{(1,1)}$ degenerates to a tri-diagonal matrix with all main diagonal entries equal to zero, i.e., ${\bf J}_{N}^{(1,1)}[n,n] = 0$, and all super- and sub-diagonal entries equal to $1/2$, i.e., ${\bf J}_{N}^{(1,1)}[n,n-1] = {\bf J}_{N}^{(1,1)}[n-1,n] = 1/2$.
We observe that this limiting Jacobi matrix is identical to the case when $\alpha =\beta= -1/2$. Thus, as $N \to \infty$, the matrix ${\bf J}_{N}^{(1,1)}$ reduces to ${\bf J}_{N}^{(-\frac{1}{2},-\frac{1}{2})}$, whose zeros are well known~\cite{szego1939orthogonal} and given by~\eqref{eq:zeroscheby}.\hfill\IEEEQEDhere}
\bibliographystyle{IEEEtran}
\bibliography{IEEEabrv,references}

\end{document}